\documentclass[9pt,shortpaper,twoside,web]{IEEEtran}
\usepackage{generic}
\usepackage{cite}
\usepackage{amsmath,amssymb,amsfonts}
\usepackage{algorithmic}
\usepackage{booktabs}
\let\labelindent\relax
\usepackage{enumitem}
\usepackage{multirow}
\usepackage{graphicx}
\usepackage{textcomp}
\usepackage[hidelinks]{hyperref}
\usepackage{xcolor}
\usepackage{xparse}
\usepackage{float}
\usepackage{scalerel}
\usepackage{tikz}
\usetikzlibrary{svg.path}

\definecolor{orcidlogocol}{HTML}{A6CE39}
\tikzset{
  orcidlogo/.pic={
    \fill[orcidlogocol] svg{M256,128c0,70.7-57.3,128-128,128C57.3,256,0,198.7,0,128C0,57.3,57.3,0,128,0C198.7,0,256,57.3,256,128z};
    \fill[white] svg{M86.3,186.2H70.9V79.1h15.4v48.4V186.2z}
                 svg{M108.9,79.1h41.6c39.6,0,57,28.3,57,53.6c0,27.5-21.5,53.6-56.8,53.6h-41.8V79.1z M124.3,172.4h24.5c34.9,0,42.9-26.5,42.9-39.7c0-21.5-13.7-39.7-43.7-39.7h-23.7V172.4z}
                 svg{M88.7,56.8c0,5.5-4.5,10.1-10.1,10.1c-5.6,0-10.1-4.6-10.1-10.1c0-5.6,4.5-10.1,10.1-10.1C84.2,46.7,88.7,51.3,88.7,56.8z};
  }
}

\newcommand\orcidid[1]{\href{https://orcid.org/#1}{\mbox{\scalerel*{
    \begin{tikzpicture}[yscale=-1,transform shape]
        \pic{orcidlogo};
    \end{tikzpicture}
}{|}}}}

\def\BibTeX{{\rm B\kern-.05em{\sc i\kern-.025em b}\kern-.08em
    T\kern-.1667em\lower.7ex\hbox{E}\kern-.125emX}}
\begin{document}
\title{MIN2Net: End-to-End Multi-Task Learning for Subject-Independent\\Motor Imagery EEG Classification}
\author{
    Phairot Autthasan$^{\orcidid{0000-0002-9566-8382}}$, \IEEEmembership{Student member, IEEE}, Rattanaphon~Chaisaen$^{\orcidid{0000-0003-1521-9956}}$, \IEEEmembership{Student member, IEEE},\\
    Thapanun Sudhawiyangkul$^{ \orcidid{0000-0003-2486-9280}}$, Phurin Rangpong, Suktipol Kiatthaveephong, Nat Dilokthanakul,\\
    Gun Bhakdisongkhram, Huy Phan$^{\orcidid{0000-0003-4096-785X}}$, \IEEEmembership{Member, IEEE}, Cuntai Guan$^{\orcidid{0000-0002-0872-3276}}$ \IEEEmembership{Fellow, IEEE},\\
    and Theerawit Wilaiprasitporn$^{\orcidid{0000-0003-4941-4354}}$, \IEEEmembership{Member, IEEE} \vspace{-0.1in}
    \thanks{This work was supported by PTT Public Company Limited, The SCB Public Company Limited, Thailand Science Research and Innovation (SRI62W1501) and Office of National Higher Education Science Research and Innovation Policy Council (C10F630057). \textit{(Phairot Autthasan and Rattanaphon Chaisaen contributed equally to this work.) (Corresponding author: Theerawit Wilaiprasitporn.)}}
    \thanks{P. Autthasan, R. Chaisaen, T. Sudhawiyangkul, P. Rangpong, S. Kiatthaveephong, N. Dilokthanakul, and T. Wilaiprasitporn are with Bio-inspired Robotics and Neural Engineering (BRAIN) Lab, School of Information Science and Technology (IST), Vidyasirimedhi\break Institute of Science \& Technology (VISTEC), Rayong, Thailand (e-mail: theerawit.w@vistec.ac.th).}
    \thanks{H. Phan is with the School of Electronic Engineering and Computer Science, Queen Mary University of London, United Kingdom.}
    \thanks{C. Guan is with the School of Computer Science and Engineering, Nanyang Technological University, Singapore.}
    \thanks{G. Bhakdisongkhram is with the School of Physical Medicine and Rehabilitation, Institute of Medicine, Suranaree University of Technology, Nakhon Ratchasima, Thailand.}
    \thanks{Code examples, and other supporting materials are available on\break\url{https://github.com/IoBT-VISTEC/MIN2Net}}
}

\maketitle

\begin{abstract}

Advances in the motor imagery (MI)-based brain-computer interfaces (BCIs) allow control of several applications by decoding neurophysiological phenomena, which are usually recorded by electroencephalography (EEG) using a non-invasive technique. Despite great advances in MI-based BCI, EEG rhythms are specific to a subject and various changes over time. These issues point to significant challenges to enhance the classification performance, especially in a subject-independent manner. To overcome these challenges, we propose MIN2Net, a novel end-to-end multi-task learning to tackle this task. We integrate deep metric learning into a multi-task autoencoder to learn a compact and discriminative latent representation from EEG and perform classification simultaneously. This approach reduces the complexity in pre-processing, results in significant performance improvement on EEG classification. \textcolor{red}{Experimental results in a subject-independent manner show that MIN2Net outperforms the state-of-the-art techniques, achieving an F1-score improvement of 6.72\%, and 2.23\% on the SMR-BCI, and OpenBMI datasets, respectively.} We demonstrate that MIN2Net improves discriminative information in the latent representation. This study indicates the possibility and practicality of using this model to develop MI-based BCI applications for new users without the need for calibration.   

\end{abstract}


\section{Introduction}
\label{sec:introduction}
\IEEEPARstart{B}{RAIN}-computer interface (BCI) systems allow users to non-muscularly communicate with a machine by classifying their neural activity patterns\cite{MCFARLAND2017194, 8945233}. Recently, electroencephalography (EEG) has been widely used as a brain-activity recording methods in BCI because it provides a non-invasive and relatively cheaper way of measuring neural activity compared to other neural acquisition techniques. Moreover, EEG offers a higher temporal resolution compared the other brain measurement techniques \cite{6737255, 6683068}.

Four main types of neurophysiological patterns are widely used to develop EEG-based BCI applications. These include steady-state visual evoked potential (SSVEP), event-related potential (ERP), movement-related cortical potentials (MRCPs), and motor imagery (MI)\cite{8926475, 6955799, 8960436, 9130151}. Among these EEG measurements, MI, used in BCI systems, has been gaining more attention because it allows users to generate the suppression of oscillatory neural activity in specific frequency bands over the motor cortex region without external stimuli \cite{PFURTSCHELLER19991842}. The neurophysiological patterns of MI originate from changing brain areas' activations in the sensorimotor cortices similar to limb movements. Furthermore, a recent study has demonstrated MI-based BCI as an assistive tool in motor rehabilitation in paralyzed patients, such as post-stroke patients\cite{7802578}.

Most MI-based BCI applications rely on a subject-dependent setting. New users have to participate in the calibration process before using a BCI system, which is time-consuming, inconvenient, and exhausting. Recently, numerous zero-calibration methods have been proposed to diminish the number of calibration trials\cite{8737742, ZHANG20211}. One prominent method is a calibration-free, or a subject-independent, where training and testing data are from different subjects. This method exhibits the ability to offer new users to use the BCI system without the calibration phase\cite{Hinton504, 8897723}. It is essential to develop reliable methods based on the subject-independent setting while preserving classification performance in an acceptable range. Thus, it is a challenge to find discriminative MI-EEG features that generalize across subjects. 


One conventional hand-crafted feature is the power spectral density (PSD). Event-related desynchronization/synchronization (ERD/ERS) is the brain activity patterns from particular frequency bands (mu (9--13 Hz) and beta (22--29 Hz)) over the sensorimotor cortex region while performing MI \cite{MCFARLAND2017194, 1506823}. Therefore we can carry out MI classification by considering PSD of EEG \cite{7802578, 939829}. However, there are some limitations of hand-crafted features. The major limitation is the selection of distinguishable information within the raw EEG (e.g., frequency bands and EEG channels), merely depending on the prior knowledge of experts\cite{JIAO2018582}. 

In recent years, the use of Deep Learning (DL) has been shown promising results and proven itself to be a successful set of models in the field of computer vision, speech recognition, and natural language processing\cite{facenet,Wang_2019_CVPR}. In contrast to hand-crafted features, DL methods can simultaneously learn complicated patterns from multiple dimensions of the data. Thus, many BCI researchers have proposed advanced DL architectures and significant improvements have been reported for EEG-MI classification. Specifically, the use of convolutional neural networks (CNNs) have widely applied in EEG-MI classification because it offers the ability to efficiently learn on both temporal and spatial features from EEG signals\cite{tonio_paper, 8310961, Lawhern_2018}. To extract the temporal and spatial connectivity patterns effectively, the combination of 2D-CNNs and long short-term memory units (LSTMs) has been adopted\cite{8745473, 8556024}. Even though existing deep learning works have been considerably successful in EEG decoding on several MI datasets, these works perform well only in the subject-dependent task, which lacks generalization capabilities on new users.

More recently, the deep learning method based multi-task autoencoder (multi-task AE) has been employed in the field of EEG based BCI because it can efficiently learn for data compression and classification tasks simultaneously\cite{7752836, 8723080}. However, to our best knowledge, few researchers have adopted the multi-task AE to learn features from EEG to address the MI classification problem because it lacks the ability to maintain discriminative patterns of the original EEG signals. To overcome this issue, we proposed \textit{MIN2Net}, a novel end-to-end neural network architecture and loss function for training multi-task AE in the MI classification task. In this way, the proposed method is able to learn latent representations that preserve discriminative information of the original EEG data by fusing deep metric learning (DML). MIN2Net is optimized by minimizing three different loss functions simultaneously: reconstruction, cross-entropy, and triplet loss functions.

The three main contributions of this paper can be summarized as follows:
\begin{itemize}
\item We proposes a novel end-to-end architecture that can effectively extract the meaningful features from EEG data without using high-complexity EEG pre-processing, resulting in an outstanding performance in the subject-independent MI classification. Futhermore, the proposed method demonstrates the excellent performance compared to state-of-the-art algorithms in the subject-independent classification over two benchmark datasets.

\item To the best of our knowledge, this is the first study proposing deep metric learning to a multi-task AE to improve the MI classification performance. The proposed method indicates the possibility to handle with discriminative information in the latent representation. 

\item Investigation via visualization of the learned latent features is carried out to interpret the proposed method's classification superiority over other state-of-the-art algorithms.
\end{itemize}

The remainder of this paper is structured as follows. Section II explains some backgrounds and related work. Section III describes the pre-processing of EEG and the structure of the proposed method. Experimental results of the proposed method are presented in Section IV and discussed in Section V. Finally, the conclusion is explained in Section VI.

\section{Related work}
In this section, we review the development of EEG-based MI classification and then conclude the limitations on the current MI-BCI research. We also describe the concepts of AE and deep metric leaning, which relate to our work. Finally, we give an overview of our proposed method. 

\subsection{EEG-based Motor Imagery Classification}
With the advance of machine learning, BCI researchers have increasingly proposed intelligent algorithms based on a subject-dependent setting to enhance the performance of EEG-based motor imagery decoding. Common Spatial Pattern (CSP) is one of the most popular and commonly used methods in MI-based BCI\cite{895946}. Features are effectively extracted via CSP can be achieved by maximizing the differences in the variances for the two classes of EEG signals. Filter Bank Common Spatial Patterns (FBCSP)\cite{4634130} is one of advanced CSP algorithms, which is based on using multiple frequency bands instead of limiting to a specific band. FBCSP has been proven to be the state-of-the-art method in EEG-based MI classification, owing to its outstanding results\cite{second_fbcsp}. After passing EEG signals through the FBCSP, the meaningful brain features are obtained and then the most discriminative features are selected using a feature selection method such as mutual information-based best individual feature (MIBIF)\cite{4634130}. In term of classifiers, many conventional algorithms such as support vector machine (SVM) and linear discriminant analysis (LDA), can be used to classify these features\cite{8897723, second_fbcsp}. \textcolor{red}{In addition, there have been a variety of algorithms extending CSP and leading to good performance\cite{5593210, 9272754, 9349966}. Some works that utilize minimum training samples have been proposed and validated on several MI datasets\cite{NECO_a_00592, 8353425}. Although these methods have outperformed state-of-the-art methods in the subject-dependent classification task, their performance still needs improvement in the subject-independent case.}

One potential direction to improve the performance of EEG-MI classification is to leverage a large-scale EEG-MI dataset using deep learning models\cite{9175874, 8897723}. In a more recent paper, Lee et al. \cite{lee_dataset} provided an OpenBMI dataset, where EEG data is measured with a large number of subjects, in multiple sessions, using the MI-BCI paradigm. Taking advantage of a large number of training samples, the OpenBMI dataset has become to be one of the benchmark EEG datasets. In the work by Kwon et al. \cite{8897723}, a subject-independent framework based on CNN architectures was proposed using spectral-spatial feature representation to improve subject-independent MI classification, resulting in a state-of-the-art performance over the OpenBMI dataset. 

\subsection{Deep Metric Learning Model}
Deep metric learning (DML) is a method based on a distance metric concept with the goal of learning representation to measure data similarity, depending on the embedding features learned from a metric learning network\cite{Ge_2018_ECCV}. Generally, similarity metric functions such as Euclidean distance, Mahalanobis distance, and cosine distance can be directly employed as the distance metric between two points. In recent years, numerous loss functions such as contrastive loss\cite{contrastive_paper}, triplet loss\cite{7298682} and quadruplet loss\cite{quadruplet_paper} are developed for DML and to enhance feature discrimination. These loss functions are used to compute similarity measure on correlated samples to enforce samples of the same class closer to each other and push samples of different classes apart from each other. Unlike other losses such as cross-entropy loss where a single sample is used to calculate the gradient, the gradient of a DML loss relies on contrastive pairs, triplets, or quadruplets of samples. Recently, DML has been applied to the field of EEG-BCI studies and achieved promising results\cite{8257016,9116935}

\subsection{Autoencoders} 
Autoencoder (AE) concept is one of the unsupervised learning algorithms introduced in 1986\cite{fisrt_ae}. AE is typically employed for data compression, denoising, dimensionality reduction, and feature extraction\cite{Hinton504, 552239, SAE_paper}. This network architecture demonstrates an ability to learn meaningful features from either unlabeled or labeled input data to create latent representation. The learned latent representation is then used to reconstruct the original input. The training objective of this network architecture is to minimize the reconstruction loss of input data. The learned latent representation appears to be more efficient when the reconstruction data is closer to the input data.    

In recent years, advanced AE architectures have been developed and adopted for EEG. Denoising sparse autoencoder (DSAE)\cite{8429921} was proposed to improve the EEG-based epileptic seizure detection. The sparsity constraint of the DSAE makes the reconstruction of the original EEG from the corrupted EEG input more efficiently. Furthermore, a compressed sensing (CS) method based on AE was proposed to handle the biopotentials and telemonitoring system\cite{7752836}. They reveal achievements in both finding the optimal data compression and classifying electrocardiogram (ECG) and EEG signals. However, most studies only focused on using AE as an unsupervised learning method to extract the salient features of an original data\cite{8429921, 8416796}. Their models were not end-to-end learning paradigms, since their classifiers need to be trained separately with the learned features and labeled data to identify the actual class.

To address the aforementioned issue, our previous work\cite{8723080} presented ERPENet, a multi-task autoencoder-based model (multi-task AE) to jointly learn multi-task deep features from both unsupervised EEG-based ERP reconstruction and supervised EEG-based ERP classification. In particular, we demonstrated that ERPENet obtained a very good performance in extracting information shared across different datasets for an event-related potential (ERP) decoding task. In this study, we propose a new architecture and a loss function for training a multi-task AE, which is capable of handling three tasks simultaneously to deal with the EEG-based MI classification, described in the following sections. 

\section{Methods}
This section first describes the three benchmark datasets. After that, we describe the design of the proposed method and discuss its loss function. Finally, we elaborate the EEG-MI classification using the proposed method in a comprehensive study. 

\subsection{Data Description}
\label{data_desciption}
We evaluated the proposed method and other baseline methods on the BCIC IV 2a\cite{BCIC_IV_2a}, SMR-BCI\cite{BCIC_SMR} and OpenBMI\cite{lee_dataset} datasets. The first two public datasets are well-known as the benchmark datasets for MI classification, offered by Graz University of Technology. The last one is the largest public MI dataset so far, provided by Korea University. The details of the databases are explained as follows:  
\begin{enumerate}[label=(\alph*)]
\item The BCI IV 2a dataset consists of 9 healthy subjects, performing left hand, right hand, feet, and tongue imagery. The EEG signals were collected using 22 Ag/AgCl electrodes at a sampling frequency of 250 Hz. The number of recorded EEG signals were 288 trials in total, obtained from two sessions on two different days (both are in an offline manner). In this paper, the classification tasks were evaluated on 20 channels from the motor cortices area ($FC_3$, $FC_1$, $FCz$, $FC_2$, $FC_4$, $C_5$, $C_3$, $C_1$, $Cz$, $C_2$, ,$C_4$, $C_6$, $CP_3$, $CP_1$, $CPz$, $CP_2$, $CP_4$, $P_1$, $Pz$, and $P_2$) and only data from right- and left-hand MI were used. Furthermore, the EEG data was downsampled from 250 to 100 Hz. The time period of the EEG-MI data was defined as the time segment between 0s and 4s after stimulus onset, resulting in a dimension of \#subjects$\times$\#trials$\times$\#channels$\times$\#sampled time points (9$\times$144$\times$20$\times$400). 
  
\item The SMR-BCI dataset contains EEG signals of 15 channels for the two-class MI task (executing the imagination of right hand and feet) recorded from 14 healthy subjects. The EEG data were gathered using a sampling frequency of 512 Hz. There were 160 trials per subject, obtained from two sessions. The first session provided 100 trials of the recorded EEG without feedback. The rest of 60 trails was the EEC recorded with feedback, acquired from the last session. Likewise, the EEG data was downsampled from 512 to 100 Hz. The four-second EEG-MI data was identified as the time interval between 0s and 4s after stimulus onset, leading to a dimension of \#subjects$\times$\#trials$\times$\#channels$\times$\#sampled time points (14$\times$160$\times$15$\times$400).

\item The OpenBMI dataset comprises 54 healthy subjects, imaging left- and right-hand movements. The recorded EEG contained 62 EEG channels with a sampling frequency of 1000 Hz and four sessions. Two sessions were defined as an offline condition because the subjects executed the MI task without feedback. The remaining sessions provided feedback for the subjects while performing the MI task, so-called on-line condition. In this paper, 20 electrodes ($FC_5$, $FC_3$, $FC_1$, $FC_2$, $FC_4$, $FC_6$, $C_5$, $C_3$, $C_1$, $Cz$, $C_2$, ,$C_4$, $C_6$, $CP_5$, $CP_3$, $CP_1$, $CPz$, $CP_2$, $CP_4$, and $CP_6$) from the motor cortices region were chosen. Furthermore, the signals were downsampled from 1,000 to 100 Hz. The time interval of EEG between 0s and 4s after stimulus onset was selected as the MI period. Finally, The segmented EEG-MI data contained a collection of \#subjects$\times$\#trials$\times$\#channels$\times$\#sampled time points (54$\times$400$\times$20$\times$400).
\end{enumerate} 

\subsection{Time-domain EEG Representation}
In this study, the raw EEG data is time-domain signals that change over time. Since the discriminative features of motor imagery are mainly distributed between 8 Hz and 30 Hz\cite{lee_dataset,8897723}, a fifth-order Butterworth band-pass filter is adopted to construct the filtered EEG data in the corresponding frequency bands. Towards this end, the filtered EEG data is used as the input of MIN2Net. Formally, we consider $x\in\mathbb{R}^{C\times T}$ as a single-trial filtered EEG data from $k$ classes, and its corresponding label is defined to be $y\in\{1,2,...,k\}$, where $C$ is the number of channels, and $T$ is the number of sampled time points.  


\begin{figure*}[t]
\centering
\includegraphics[width=1.5\columnwidth]{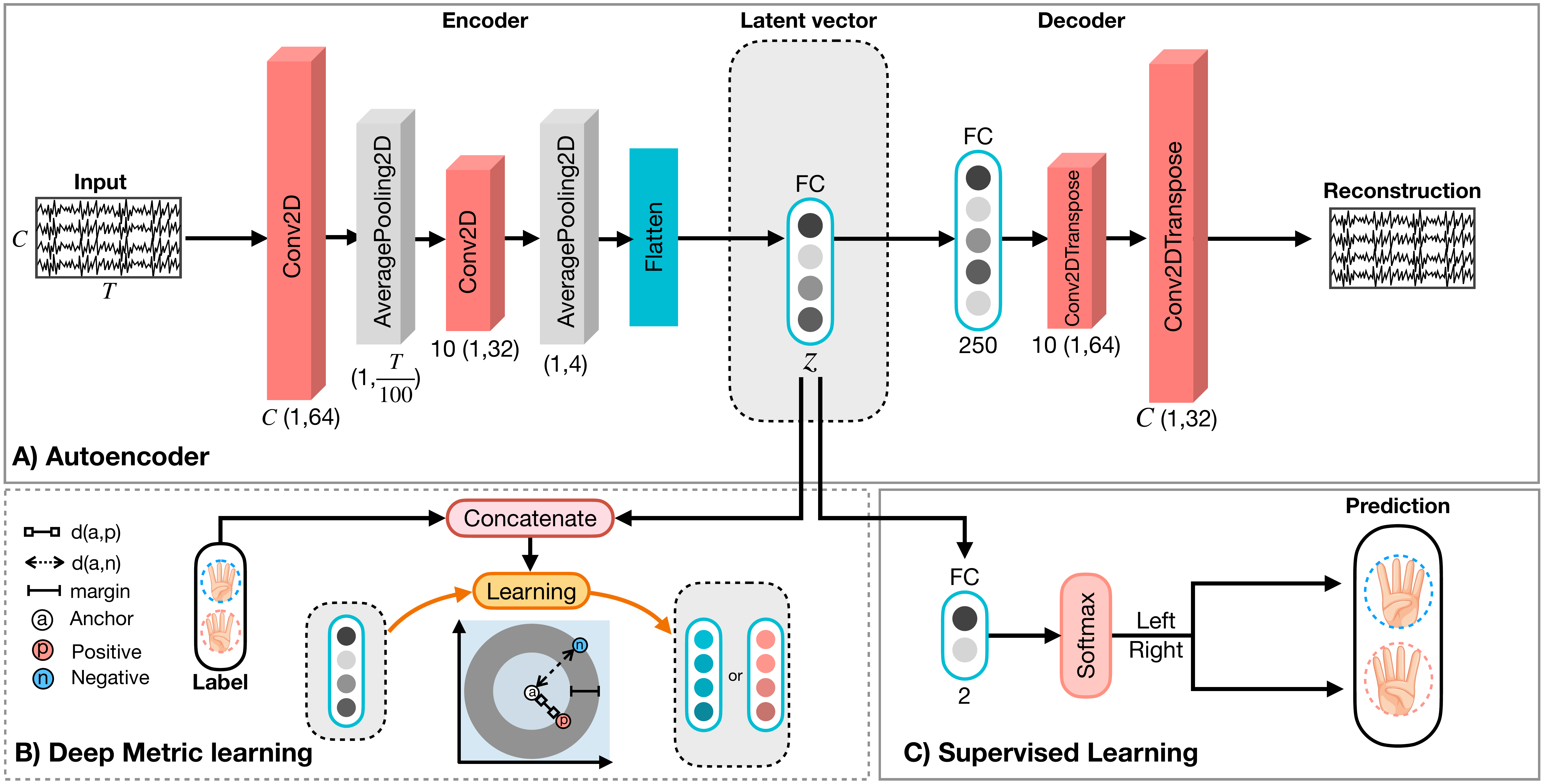}
\caption{Overall visualization of MIN2Net Architecture. (a) exhibits the AE network consisting of 3 components: encoder, latent vector, and decoder. The encoder compresses the input data and produces the latent vector then the decoder reconstructs the input data from this latent vector. (b) illustrates metric learning that learns to minimize the distances of embedding vectors of the same label while maximizing different labels. (c) displays the supervised classifier---the latent vector was fed into a FC layer using softmax activation for classification. Full details about the
network architecture can be found in \autoref{tab:model}.}
\label{fig:model}
\end{figure*}

\begin{table*}
\caption{MIN2Net architecture, where \textit{C} is the number of channels, \textit{T} is the number of time points, \textit{z} is the size of latent vector and \textit{N} is the number of classes. Noted that the data format of Conv2D is ``channels last''}
\label{tab:model}
\centering
\resizebox{1.35\columnwidth}{!}{%
\begin{tabular}{@{}lllllllll@{}}
\toprule[0.2em]
\multicolumn{2}{l}{\textbf{Blocks}}                      & \multicolumn{1}{l}{\textbf{Layer}} & \textbf{Filter} & \multicolumn{1}{l}{\textbf{Size}} & \multicolumn{1}{l}{\textbf{Stride}} & \multicolumn{1}{l}{\textbf{Activation}} & \multicolumn{1}{l}{\textbf{Options}} & \multicolumn{1}{l}{\textbf{Output}} \\ \midrule[0.1em] 
\multirow{13}{*}{Autoencoder} & \multirow{8}{*}{Encoder} & Input                              &                 &              (1,\textit{T},\textit{C})                      &                                     &                                         &                                      & (1,\textit{T},\textit{C})                            \\
                              &                          & Conv2D                             & \textit{C}               & (1,64)                            & 1             & ELU                                     & padding=same                         & (1,\textit{T},\textit{C})                            \\
                              &                          & BatchNormalization                          &                 &                                   &                                     &                                         &                                      & (1,\textit{T},\textit{C})                            \\
                              &                          & AveragePooling2D                      &                 & (1, \textit{T}//100)                       &                                     &                                         &                                      & (1,100,\textit{C})                           \\
                              &                          & Conv2D                             & 10              & (1,32)                            & 1              & ELU                                     & padding=same                         & (1,100,10)                          \\
                              &                          & BatchNormalization                                 &                 &                                   &                                     &                                         &                                      & (1,100,10)                          \\
                              &                          & AveragePooling2D                   &                 & (1,4)                             &                                     &                                         &                                      & (1,25,10)                           \\
                              &                          & Flatten                            &                 &                                   &                                     &                                         &                                      & (250)                               \\ \cmidrule[0.1em](l){2-9} 
                              & Latent                   & FC                                 &                 & (\textit{z})                               &                                     &                                         &                                      & (\textit{z})                                 \\ \cmidrule[0.1em](l){2-9} 
                              & \multirow{4}{*}{Decoder} & FC                                 &                 & (250)                             &                                     &                                         &                                      & (250)                               \\
                              &                          & Reshape                            &                 & (1, 25, 10)                       &                                     &                                         &                                      & (1, 25, 10)                         \\
                              &                          & Conv2DTranspose                    & 10              & (1,64)                            & 4               & ELU                                     & padding=same                         & (1, 100, 10)                        \\
                              &                          & Conv2DTranspose                    & \textit{C}               & (1,32)                            & \textit{T}//100                              & ELU                                     & padding=same                         & (1,\textit{T},\textit{C})                            \\ \midrule[0.1em] 
\multicolumn{2}{c}{\multirow{3}{*}{Metric learning}}     & Input                              &                 & (1)                               &                                     &                                         &                                      & (1)                                 \\
\multicolumn{2}{c}{}                                     & Latent                          &                 &                                  &                                     &                                         &                                      & (\textit{C})                                 \\
\multicolumn{2}{c}{}                                     & Concatenate                        &                 &                                   &                                     &                                         & {[}Input,Latent{]}                & (\textit{C}+1)                               \\ \midrule[0.1em]
\multicolumn{2}{c}{\multirow{2}{*}{Supervised Learning}}          & Latent                          &                 &                                   &                                     &                                         &                                      & (\textit{C})                                 \\
\multicolumn{2}{c}{}                                     & FC                                 & \textit{N}             &                                   &                                     & softmax                                 &                                      & (\textit{N})                                 \\ \bottomrule[0.2em]
\multicolumn{9}{l}{\textit{z}  is equal to \textit{C} and 256 for 2- and 3-class classification, respectively.} \\
\end{tabular}
}
\end{table*}

\subsection{Proposed Architecture: MIN2Net}
An overview of our proposed MIN2Net is illustrated in \autoref{fig:model} and the detail configuration of each layer is shown in \autoref{tab:model}. The MIN2Net is composed of three main modules: autoencoder, deep metric learning, and supervised learning.

\subsubsection{Autoencoder}
The autoencoder module in the MIN2Net consists of two major components the encoder $z=q(x)$ and the decoder $\hat{x}=p(z)$ components. In the encoder component, an input signal $x$ is encoded into a latent vector $z$ by reducing the input signal's dimension. For the decoder component, the given latent vector $z$ is decoded back to the input signal $\hat{x}$.

The encoder has two CNN blocks, each of them consists of a Conv2D layer, a batch normalization (BN) layer, an exponential linear unit (ELU), and an average pooling layer (AveragePooling2D). The final CNN layer's output is considered as the input of a fully connected layer for mapping the latent representation.  

Inspired by CSP, this study utilizes the CNN approach as spatial filtering to effectively learn discriminative features from a set of EEG inputs ($x$). Each CNN block is operated on the channel mixing CNN concept\cite{8310961}, combining all channels of the input signals. The convolution operation is performed based on the linear combination of all the given channels, convoluted along the time dimension. Therefore, the output is constructed as a new time-series signal, simultaneously extracting spatial information from all the feature channels. Here, the encoder's hidden size is large for the first CNN layer but is gradually decreased in the following CNN layers. More details of the layers' parameters are shown in \autoref{tab:model}. The average pooling layers are applied to extract the important features of the given input signals and reduce the number of parameters. The main benefit of applying the average pooling is to exploit layers with local filters to share weights among all channels of the given input signals. After every CNN layer, a BN layer is used before feeding into the subsequent average pooling layer. The feature maps after the final average pooling are transformed into a vector representation via flattening. Finally, the flattened vector is presented to a fully connected (FC) layer with the hidden size of $z$ units to embed and produce the latent vector. The latent vector size of $z$ is expected to preserve the robust features for EEG-MI signals.  

For the decoder component, the decoder structure is arranged in a symmetrical way to the encoder component. Since it is essential to match the CNN blocks' input dimension, the latent vector is passed through a FC layer and then fed into a reshape layer to construct the data in a suitable dimensions. Each of the two CNN blocks of the decoder component makes use of a transpose convolution layer (Conv2DTranspose) with a stride of 4 and an ELU layer. A stride of 4 is employed to upsample the data's size in a similar way to an upsampling layer. The transpose convolution can extract useful features and reduce useless features, which is beneficial for reconstructing the latent vector. Consequently, the constructed input signal is obtained after passing the two CNN blocks.

The training objective of the AE module is to minimize the reconstruction error between the input and the reconstruction. Here, we employ the mean square error (MSE) as the loss function. Given the input signals $x_j= \{x_1,x_2,...,x_C\}$, the loss function is expressed as:
\begin{equation} \label{eq_mse}
\mathcal{L}_{\textrm{MSE}}(x,\hat{x}) = \frac{1}{C} \sum_{j=1}^{C} \| x_j - \hat{x_j} \|^2.
\end{equation}
Where $\hat{x_j}$ is the reconstruction signal of the channel $j$.

\subsubsection{Deep Metric Learning}
To preserve the distinguishable patterns in the latent representation of the AE, a deep metric leaning module (DML) was introduced in the AE, extended from the latent vector. With the DML, the network is tasked to learn a distance metric via which the discrimination of the learned features is enhanced. In this paper, we employ a triplet loss in the DML module to reflect the relative distances among different classes of the latent vectors. Owing to avoiding a risk of slow convergence and pool local optima, we decide to use a semi-hard triplet constraint, which was demonstrated good performance in \cite{Schroff_2015}. During training, a set of triplets $\{x^a , x^p , x^n\}$ is randomly sampled from the training data, where the anchor sample $x^a$ is closer to the positive sample $x^p$ than the negative sample $x^n$. Subsequently, the triplet of three input signals are passed through the encoder component concurrently to obtain their latent vector $z^a$ , $z^p$ and $z^n$. Thus, the loss function can be formulated as:
\begin{equation} \label{eq_triplet_loss}
\mathcal{L}_{\textrm{triplet}}(z^a,z^p,z^n) =  \frac{1}{2}\left[\|z^a - z^p\|^2 - \|z^a - z^n\|^2 + \alpha\right]_+.
\end{equation}
where $[z]_+ = max(z,0)$. The threshold $\alpha$ is the margin parameter that enforces the Euclidean distance $\|z^a - z^p\|^2$ of positive pairs to be shorter than the Euclidean distance $\|z^a - z^n\|^2$ of negative pairs. Importantly, the margin of the triplet loss plays a significant role in training the DML module.

\begin{figure*}
\centering
\includegraphics[width=1.5\columnwidth]{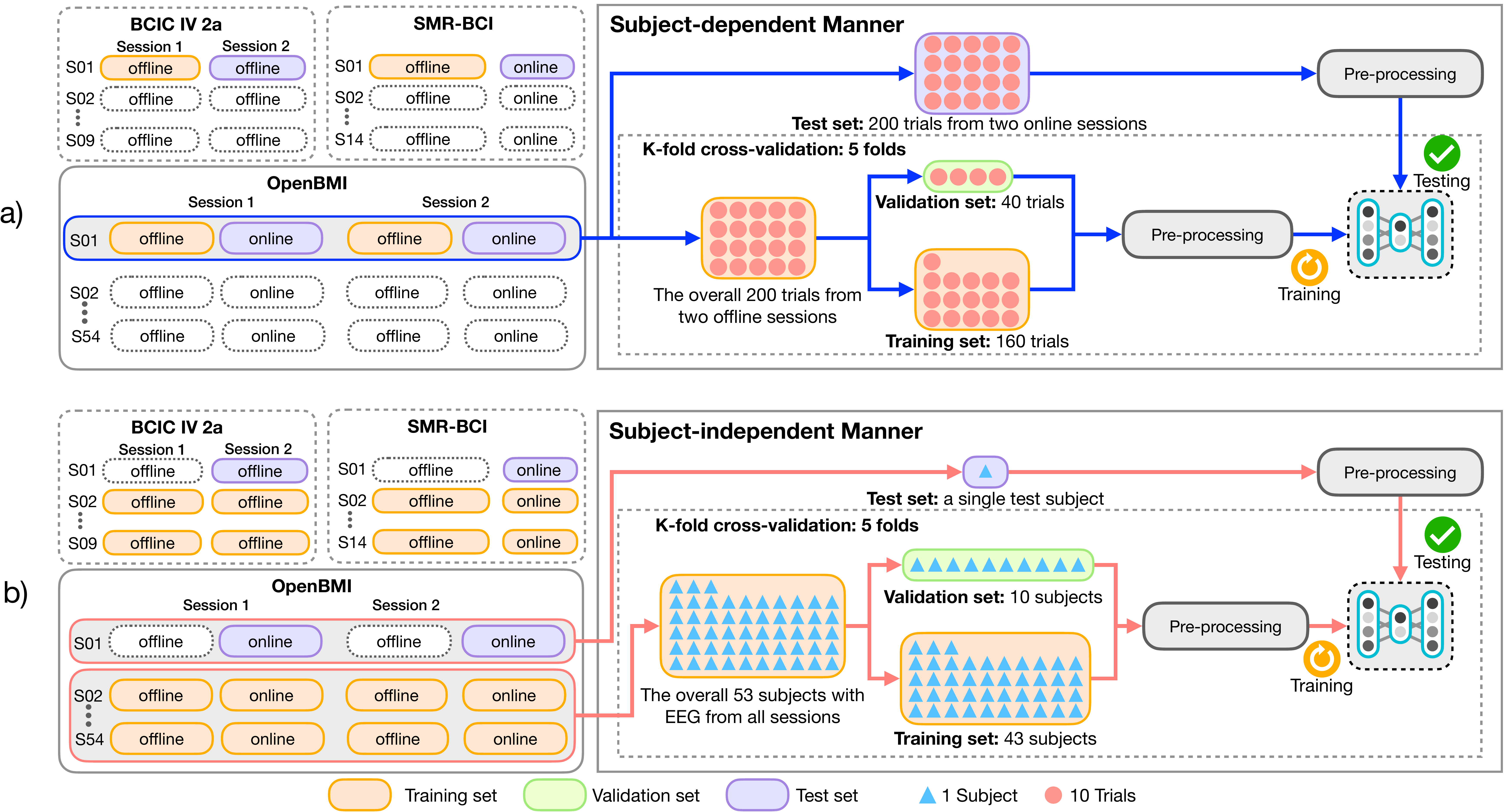}
\caption{Framework of a) subject-dependent and b) subject-independent with stratified k-fold cross-validation for the classification models.}
\label{fig:framework}
\end{figure*}

\subsubsection{Supervised Learning}
This module utilizes a standard softmax classifier as a supervised classifier to classify the underlying latent vectors of the input EEG signals. The latent vector $z$ is fed into the FC layer with the softmax activation to obtain the weight of importance for each class, expressed as follows: 
\begin{equation} \label{fc_softmax}
\hat{y}(z) = softmax(Wz+b)
\end{equation}
Where $W$ and $b$ are the weight matrix and the bias vector respectively. Then, the model is trained using Adam optimizer to minimize the cross-entropy loss, calculated as:
\begin{equation} \label{eq_cross-entropy}
\mathcal{L}_{\textrm{cross-entropy}}(y,\hat{y}) = -\sum_{k=1}^{\mid class \mid} y_k\log{\hat{y}_k}.
\end{equation}
where $y$ and $\hat{y}$ are the true label and the classification probabilities respectively. The class with the maximum classification probability is identified as the predicted class of the single-trial EEG signal.

\subsection{Training Procedure for MIN2Net}
\label{sec:loss}
The training objective of the proposed method is optimized by incorporating the three loss functions: $\mathcal{L}_{\textrm{MSE}}$ in \autoref{eq_mse}, $\mathcal{L}_{\textrm{triplet}}$ in \autoref{eq_triplet_loss}, and $\mathcal{L}_{\textrm{cross-entropy}}$ in \autoref{eq_cross-entropy}. The final loss function of our MIN2Net model $\mathcal{L}_{\textrm{MIN2Net}}$ is expressed as:
\begin{equation} \label{eq_total_loss}
\begin{aligned}
\mathcal{L}_{\textrm{MIN2Net}}(x,\hat{x},z^a,z^p,z^n,y,\hat{y}) =\frac{1}{N}\sum_{i=1}^{N}\{\beta_1\mathcal{L}_{\textrm{MSE}}(x_i,\hat{x_i})\\
+\beta_2\mathcal{L}_{\textrm{triplet}}(z_i^a,z_i^p,z_i^n)+\beta_3 \mathcal{L}_{\textrm{cross-entropy}}(y_i,\hat{y_i})\}. 
\end{aligned}
\end{equation}
Where $N$ denotes the total number of input signals. $\beta_1$, $\beta_2$ and $\beta_3$ represent the hyperparameters to weight the contribution of each loss function. As a result of integrating the three loss functions, both the unsupervised and the DML are able to influence the learning process when the supervised learning occurs.

\subsection{Network Training}
The proposed method was implemented using the Keras framework (TensorFlow v2.2.0 as backend). The training process was implemented using NVIDIA Tesla v100 GPU with 32GB memory. \textcolor{red}{In each training iteration, the loss function was optimized by utilizing Adam optimizer with the learning rate schedule between [$10^{-3}$, $10^{-4}$] for the binary classification task and the learning rate schedule between [$10^{-4}$, $10^{-5}$] for the multi-class classification task.} The learning rate of two considered tasks was decreased with a decay rate of 0.5 when there was no improvement in the validation loss for 5 consecutive epochs. We set a batch size of 10 samples for the subject-dependent classification setting and 100 samples for the subject-independent classification setting. Finally, the number of training iterations relied on the early stopping strategy so that the training process was stopped if there was no reduction of the validation loss for 20 consecutive epochs.

\subsection{Baseline Methods}
To demonstrate the effectiveness of our MIN2Net, we implemented four state-of-the-art methods for comparison. All deep learning approaches were implemented using the Keras framework (TensorFlow v2.2.0 as backend). 
\subsubsection{FBCSP-SVM}
FBCSP was developed upon the idea of the original CSP algorithm\cite{4634130}. By using the FBCSP as the feature extraction method, the distinguishable EEG features were extracted from multiple frequency bands. In this paper, the FBCSP was implemented using MNE-Python package (version 0.20)\cite{mne_package} and then applied with 4 spatial filters to decompose EEG signals into 9 frequency bands with a bandwidth of 4 Hz from 4 to 40 Hz (4--8 Hz, 8--12 Hz, ..., 36--40 Hz). Here, each frequency band was created using bandpass filtering with 5\textsuperscript{th} order non-causal Butterworth filter. Subsequently, a support vector machine (SVM) was used to classify MI by incorporating a grid search algorithm. For the SVM classifier, the hyperparameters consisted of kernel (linear, radial bias function (RBF), sigmoid), $C$ (0.001, 0.01, 0.1, 1, 10, 100, 1000), and, particularly for RBF kernel, gamma (0.01, 0.001). With respect to the grid search algorithm, the prediction on the validation set of the classification was assessed to obtain the optimal set of hyperparameters. Eventually, the SVM classifier with the optimal parameters was used for testing purpose.

\subsubsection{Deep Convnet}
Deep Convnet was introduced as a DL model based on two CNN architectures\cite{tonio_paper} and proven to be effective for dealing with EEG-MI classification. In this study, the Deep Convnet was implemented in the optimal parameters as done in \cite{tonio_paper}. Moreover, the raw EEG data was band-pass filtered between 8 and 30 Hz (5\textsuperscript{th} order non-causal Butterworth filter).   

\subsubsection{EEGNet-8,2}
Inspired by the FBCSP method, EEGNet-8,2 was proposed as a compact CNN architecture to capture discriminative EEG features, which achieved a very good performance in different BCI paradigms\cite{Lawhern_2018}. Here, EEGNet-8,2 was reproduced to offer a comparable performance. The network parameters were kept in the optimal set of hyperparameters as recommended in the original publication\cite{Lawhern_2018}. Furthermore, the raw EEG data were likewise pre-processed with the same protocol as in the Deep Convnet model to construct the input for the training of EEGNet-8,2.  

\subsubsection{Spectral-spatial CNN}
The spectral-spatial CNN framework based on CNN architectures (spectral-spatial CNN) was presented by\cite{8897723} and demonstrated state-of-the-art performance in a subject–independent MI decoding. The framework learned the spectral-spatial input, capturing discriminative features from the EEG signals' multiple frequency bands. In this paper, the raw EEG signals were similarly constructed the spectral-spatial representation as done in \cite{8897723}. The spectral-spatial CNN model was implemented in the optimal parameters as defined in the original paper.

\begin{table}
\caption{List of the optimal set of hyperparameters for MIN2Net}
\label{tab:beta_param}
\centering
\resizebox{0.7\columnwidth}{!}{%
    \begin{tabular}{@{}ccccccc@{}}
    \toprule[0.2em]
    \multirow{3}{*}{\textbf{Dataset}} & \multicolumn{6}{c}{\textbf{Parameter}}                                                   \\ \cmidrule[0.1em](l){2-7} 
                             & \multicolumn{3}{c}{\textbf{Subject-dependent}} & \multicolumn{3}{c}{\textbf{Subject-independent}} \\ \cmidrule[0.1em](l){2-4} \cmidrule[0.1em](l){5-7}
                             & \textbf{$\beta_1$}          & \textbf{$\beta_2$}         & \textbf{$\beta_3$}         & \textbf{$\beta_1$}          & \textbf{$\beta_2$}          & \textbf{$\beta_3$}          \\ \midrule[0.1em]
    BCIC IV 2a        & 1.0         & 0.1         & 1.0     & 0.5         & 0.1        & 1.0  \\ 
    SMR-BCI           & 0.1         & 0.1         & 1.0     & 0.1         & 1.0        & 0.1   \\ 
    OpenBMI           & 0.5         & 0.5         & 1.0     & 0.5         & 0.5        & 1.0       \\ \bottomrule[0.2em]
    \end{tabular}%
    }
\end{table}

\subsection{Experimental Evaluation}
To demonstrate our MIN2Net method as a generalized MI classification model, we conducted the experiments with both subject-dependent and subject-independent manners on three benchmark datasets (BCIC IV 2a, SMR-BCI, and OpenBMI). 
Accuracy and F1-score were used to evaluate the performance of all considered methods.

\autoref{fig:framework}(a) exhibits an example of how we divided the training and testing sets in a subject-dependent manner. For BCI IV 2a dataset, the offline data from session 1 was used as the training set, and the offline from session 2 was used as the testing set. Meanwhile, the training and testing sets were obtained from the offline and online sessions for both SMR-BCI and OpenBMI datasets. \textcolor{red}{The stratified 5-fold CV was then utilized to split the training set into the new training and validation sets for parameter search. Each fold was performed by preserving 50 percent of samples for each class in the new training and validation sets.}

The subject-independent manner was conducted with a leave-one(subject)-out cross-validation (LOSO-CV), as illustrated in \autoref{fig:framework}(b). Suppose that there are $N_s$ subjects for a specific dataset. In each fold of LOSO-CV, a single subject was used as the testing set, and the remaining $N_s-1$ subjects were employed as the training set to obtain $N_s$ classification results. The training set was constructed using all data sessions of all $N_s-1$ training subjects for each dataset. Meanwhile, we chose the offline session 2 of the test subject as the test set of BCI IV 2a dataset, and the online session of the test subject as the test set for both SMR-BCI and OpenBMI datasets. Furthermore, the stratified 5-fold CV scheme was adopted on the training set to find an optimal set of all classifier parameters. Finally, we calculated the average classification accuracy from the $N_s\times5$ evaluations as the overall performance of MIN2Net and other baseline methods.   

The details of the four experiments of the entire study were described as follows:
\subsubsection{Experiment I}
To find the optimal set of hyperparameters of MIN2Net, we conducted initial experiments for parameter search. We first experimented with EEG-MI binary classification to tune $\beta$ parameters for the loss function of MIN2Net in \autoref{eq_total_loss}. The grid search algorithm was carried out in the set of $\{0.1, 0.5, 1.0\}$ for $\beta_1$, $\beta_2$, and $\beta_3$. As shown in \autoref{eq_triplet_loss}, the margin $\alpha$ plays a significant role during training MIN2Net. To examine the effect of this hyperparameter, we performed experiments with MIN2Net to search for an optimal value of $\alpha$ in the set $\{0.1, 0.5, 1.0, 5.0, 10.0, 100.0\}$.

\textcolor{red}{Furthermore, we conducted an ablation study to examine the effectiveness of each component in MIN2Net. We compared our complete MIN2Net model with two modification models:
\begin{itemize}
  \item MIN2Net-without triplet: the MIN2Net without the DML module
  \item MIN2Net-without decoder: the MIN2Net without decoder part of AE module
\end{itemize}
}
\textcolor{red}{We then performed one-way repeated measures analysis of variance (ANOVA) with Bonferroni correction to evaluate the significant differences among the classification performance of MIN2Net and the aforementioned modification models.
}


\begin{table}
    \caption{Classification performance (Accuracy $\pm$ SD and F1-score $\pm$ SD) in \% of MIN2Net using the subject-dependent and subject-independent manners comparisons on six different margins ($\alpha$). Bold denotes the best numerical values.}
    \label{tab:AE_triplet_margin_results}
    \centering
    \resizebox{1.0\columnwidth}{!}{%
        \begin{tabular}{@{}cccccc@{}}
        \toprule[0.2em]
        \multirow{2}{*}{\textbf{Dataset}} & \multirow{2}{*}{\begin{tabular}[c]{@{}c@{}}\textbf{Margin} \\ \textbf{($\alpha$)}\end{tabular}} & \multicolumn{2}{c}{\textbf{Subject-dependent}} & \multicolumn{2}{c}{\textbf{Subject-independent}} \\  \cmidrule[0.1em](l){3-4}\cmidrule[0.1em](l){5-6} 
                                          &                                            & \textbf{Accuracy}      & \textbf{F1-score}     & \textbf{Accuracy}       & \textbf{F1-score}      \\ \midrule[0.1em]
        
        \multirow{6}{*}{\begin{tabular}[c]{@{}c@{}}BCIC \\ IV 2a\end{tabular}}       & 0.1                              & 62.28 $\pm$ 13.90          & 62.61 $\pm$ 15.13          & 58.64 $\pm$ 8.57           & 46.59 $\pm$ 23.58          \\
                                          & 0.5                              & 62.25 $\pm$ 13.89          & 63.08 $\pm$ 14.70          & 59.27 $\pm$ 8.38           & 49.01 $\pm$ 19.29          \\
                                          & 1.0                                & 63.46 $\pm$ 14.33          & 64.28 $\pm$ 15.27          & \textbf{60.03 $\pm$ 9.24}  & \textbf{49.09 $\pm$ 23.28} \\
                                          & 5.0                                & 63.66 $\pm$ 13.65          & 64.37 $\pm$ 14.57          & 59.61 $\pm$ 8.84           & 49.53 $\pm$ 19.56          \\
                                          & 10.0                               & 63.87 $\pm$ 14.51          & 64.03 $\pm$ 15.66          & 59.85 $\pm$ 8.44           & 48.39 $\pm$ 20.37          \\
                                          & 100.0                              & \textbf{65.23 $\pm$ 16.14} & \textbf{64.72 $\pm$ 18.39} & 58.78 $\pm$ 8.69           & 46.14 $\pm$ 24.29          \\ \midrule[0.1em]
       \multirow{6}{*}{SMR-BCI}           & 0.1                              & 64.00 $\pm$ 15.51            & 62.47 $\pm$ 16.60            & 56.76 $\pm$ 11.19            & 57.83 $\pm$ 20.50            \\
                                          & 0.5                              & 64.31 $\pm$ 15.70            & 62.27 $\pm$ 17.07            & 58.45 $\pm$ 12.67            & 58.41 $\pm$ 22.40            \\
                                          & 1.0                                & \textbf{65.90 $\pm$ 16.50} & \textbf{64.13 $\pm$ 17.66} & \textbf{59.79 $\pm$ 13.72} & \textbf{61.10 $\pm$ 23.64} \\
                                          & 5.0                                & 65.14 $\pm$ 16.08            & 62.04 $\pm$ 18.20            & 59.69 $\pm$ 13.86            & 58.88 $\pm$ 22.48            \\
                                          & 10.0                               & 65.45 $\pm$ 15.81            & 62.01 $\pm$ 17.82            & 58.81 $\pm$ 13.50            & 58.61 $\pm$ 23.43            \\
                                          & 100.0                              & 66.98 $\pm$ 17.22            & 62.77 $\pm$ 20.58            & 60.79 $\pm$ 13.73            & 60.47 $\pm$ 24.31            \\ \midrule[0.1em]
       \multirow{6}{*}{OpenBMI}           & 0.1                              & 59.25 $\pm$ 14.27            & 61.79 $\pm$ 14.26            & 72.14 $\pm$ 14.22            & 72.07 $\pm$ 15.19            \\
                                          & 0.5                              & 59.85 $\pm$ 13.93            & 62.17 $\pm$ 14.17            & 71.06 $\pm$ 13.91            & 71.23 $\pm$ 14.40            \\
                                          & 1.0                                & \textbf{61.03 $\pm$ 14.47} & \textbf{63.59 $\pm$ 14.52} & \textbf{72.03 $\pm$ 14.04} & \textbf{72.62 $\pm$ 14.14} \\
                                          & 5.0                                & 59.93 $\pm$ 13.77            & 62.41 $\pm$ 14.31            & 70.43 $\pm$ 13.81            & 71.00 $\pm$ 13.90            \\
                                          & 10.0                               & 58.97 $\pm$ 13.56            & 61.51 $\pm$ 14.14            & 69.43 ±14.29             & 69.46 $\pm$ 15.32            \\
                                          & 100.0                              & 57.14 $\pm$ 13.96            & 59.11 $\pm$ 15.98            & 70.48 $\pm$ 14.40            & 70.95 $\pm$ 15.25            \\  \bottomrule[0.2em]
        \end{tabular}
    }
    \end{table}

\begin{table*}
\caption{Classification performance (Accuracy $\pm$ SD and F1-score $\pm$ SD) in \% of MIN2Net compared to MIN2Net-without triplet and MIN2Net-without decoder using the subject-dependent and subject-independent manners on all datasets. Bold denotes the best numerical values, and * represents the performance value which was significantly higher than all comparison pairs, $p < 0.05$.}
\label{tab:AE_wo_triplet}
\centering
\resizebox{1.35\columnwidth}{!}{%
    \begin{tabular}{@{}cccccc@{}}
    \toprule[0.2em]
    \multirow{2}{*}{\textbf{Dataset}} & \multirow{2}{*}{\textbf{Comparison Model}} & \multicolumn{2}{c}{\textbf{Subject-dependent}} & \multicolumn{2}{c}{\textbf{Subject-independent}} \\  \cmidrule[0.1em](l){3-4}\cmidrule[0.1em](l){5-6} 
                                      &                                            & \textbf{Accuracy}      & \textbf{F1-score}     & \textbf{Accuracy}       & \textbf{F1-score}      \\ \midrule[0.1em]
    
\multirow{3}{*}{BCIC IV 2a}         & MIN2Net-without triplet          & 60.76 $\pm$ 11.93                                  & 61.09 $\pm$ 13.83                                 & 58.70 $\pm$ 8.91                            & \textbf{49.36 $\pm$ 20.25}                         \\
                                    & MIN2Net-without decoder       & \textbf{65.71 $\pm$ 16.16}                                  & \textbf{65.46 $\pm$ 18.34}                                 & 57.55 $\pm$ 9.06                            & 44.24 $\pm$ 24.89                          \\
                                    & MIN2Net                          & 65.23 $\pm$ 16.14                                  & 64.72 $\pm$ 18.39                                 & \textbf{60.03 $\pm$ 9.24}                            & 49.09 $\pm$ 23.28                          \\ \midrule[0.1em]
\multirow{3}{*}{SMR-BCI}            & MIN2Net-without triplet          & 63.86 $\pm$ 14.13                                  & 61.31 $\pm$ 16.19                                 & 57.95 $\pm$ 12.55                           & 60.53 $\pm$ 20.33                          \\
                                    & MIN2Net-without decoder       & 64.86 $\pm$ 16.21                                  & 62.65 $\pm$ 17.99                                 & 57.38 $\pm$ 12.22                           & 55.28 $\pm$ 22.17                           \\
                                    & MIN2Net                          & \textbf{65.90 $\pm$ 16.50}                                  & \textbf{64.13 $\pm$ 17.60}                                 & \textbf{59.79 $\pm$ 13.72}                           & \textbf{61.10 $\pm$ 23.64}                           \\ \midrule[0.1em]
\multirow{3}{*}{OpenBMI}            & MIN2Net-without triplet          & 59.66 $\pm$ 14.02                                  & 61.64 $\pm$ 14.44                                 & 71.10 $\pm$ 13.58                           & 69.28 $\pm$ 16.10                            \\
                                    & MIN2Net-without decoder       & 58.76 $\pm$ 13.79                                  & 61.70 $\pm$ 13.64                                 & 70.59 $\pm$ 14.23                           & 70.69 $\pm$ 14.48                             \\
                                    & MIN2Net                          & \textbf{61.03 $\pm$ 14.47*}                                  & \textbf{63.59 $\pm$ 14.52*}                                 & \textbf{72.03 $\pm$ 14.04*}                           & \textbf{72.62 $\pm$ 14.14*}                               \\  \bottomrule[0.2em]
    \end{tabular}
}
\end{table*}

\subsubsection{Experiment II} 
In this study, we compared the EEG-MI classification performance of MIN2Net with the aforementioned baseline methods. This experiment was conducted based on EEG-MI binary classification to investigate all methods' effectiveness over the three aforementioned benchmark datasets in both the subject-dependent and subject-independent scenarios. To make a fair comparison, all methods were evaluated on the same training, validation, and testing sets. We carried out one-way repeated measures analysis of variance (ANOVA) with Bonferroni correction to analyze the classification performance's significant differences between our MIN2Net and all baseline methods.

\subsubsection{Experiment III}
To demonstrate the practicality of MIN2Net in developing real-world applications, we conducted an experiment with three-class EEG-MI classification task (right hand MI vs. left hand MI vs. resting EEG) over the OpenBMI dataset. This experiment was designed to compare the effectiveness of MIN2Net against the aforementioned baseline methods in pseudo-online situations. Since MIN2Net on the subject-independent scenario demonstrated prominent performance from Experiment I and II, we used this scenario in this experiment. Here, both the right and left hand MI signals were likewise segmented with the same protocol as in \autoref{data_desciption}(c). Meanwhile, we chose the time interval of EEG between 4s and 8s after stimulus onset as the resting EEG period. Since the resting EEG was obtained from all EEG recordings, the number of trials in each MI class was less than the resting EEG class. We decided to randomly select half of all resting EEG trials to overcome the imbalance issue. To compare all the used methods, an ANOVA with Bonferroni correction was employed for statistical analysis.

\subsection{Visualization}
To compare the abilities of deep learning methods in extraction of the highly discriminative features from the EEG signals, we used the $t$-SNE method\cite{vandermaaten08a} to visualize the generalized brain features learned by different deep learning methods in the 2-dimensional embedding space. By applying the $t$-SNE, the high dimensional embedding space at the input of the final fully connected layer in all trained model was used as the learned features for visualization.

\section{Results}
This section reports the results and statistical analysis of Experiment I, II, and III to validate the effectiveness of MIN2Net. Furthermore, we visualize of the learned EEG features, to demonstrate the discriminative power of the features learned by the MIN2Net. The performance of each experiment was reported as accuracy and F1-score with standard deviation (Accuracy $\pm$ SD and F1-score $\pm$ SD).

 \subsection{Experiment I: Parameters Adjustment}
\autoref{tab:beta_param} presents the optimal hyperparameters to adjust each module's weight ($\beta_1$, $\beta_2$ and $\beta_3$) for the loss function of MIN2Net in \autoref{eq_total_loss}, resulting in the optimal performance of classifying EEG-MI data. Note that, the average classification performance of MIN2Net on all $\beta$ combinations of each dataset is reported in the supplementary materials\footnote{\url{https://github.com/IoBT-VISTEC/MIN2Net}}. \autoref{tab:AE_triplet_margin_results} illustrates the classification results when different value of the margin parameters ($\alpha$) are contributed in the DML module of MIN2Net. We observed that the margin parameter's size had a significant impact on the final classification performance, and when marking the margin value as 1.0, the MIN2Net achieved the best performance in the subject-independent manner for all datasets. For the subject-dependent manner, the margin value of 100.0 contributed to the best performance of MIN2Net on BCIC IV 2a dataset, whereas the best performance of MIN2Net on both SMR-BCI and OpenBMI datasets were obtained by setting the margin value as 1.0.          

\textcolor{red}{The results of the ablation study are summarized in \autoref{tab:AE_wo_triplet}. It can be seen that the classification performance of MIN2Net outperformed both MIN2Net-without triplet and MIN2Net-without decoder models in terms of accuracy and F1-score on both subject-dependent and subject-independent manners for both SMR-BCI and OpenBMI datasets. In a paired $t$-test, MIN2Net was significantly higher than these modification models in both manners for the OpenBMI dataset, $p<0.05$. However, on the BCIC IV 2a dataset, the performance improvement of MIN2Net to its modification models was not found in both manners.}

\begin{table*}
    \caption{Classification performance (Accuracy $\pm$ SD and F1-score $\pm$ SD) in \% for the subject-dependent and subject-independent schemes on BCIC IV 2a, SMR-BCI, and OpenBMI compared to six different methods. Bold denotes the best numerical values, and * represents the performance value which was significantly higher than all comparison pairs, $p < 0.05$.}
    \label{tab:result_all}
    \centering
    \resizebox{1.5\columnwidth}{!}{%
        \begin{tabular}{@{}ccccccc@{}}
            \toprule[0.2em]
            \multirow{2}{*}{\textbf{Datasets}}                                                                       & \multirow{2}{*}{\textbf{Comparison Model}} & \multirow{2}{*}{\textbf{End-to-end}} & \multicolumn{2}{c}{\textbf{Subject-dependent}}             & \multicolumn{2}{c}{\textbf{Subject-independent}}  \\ \cmidrule[0.1em](l){4-7} 
                                                                                                                     &                                            &                                      & \textbf{Accuracy}      & \textbf{F1-score}                 & \textbf{Accuracy}       & \textbf{F1-score}       \\ \midrule[0.1em]
            \multirow{5}{*}{\begin{tabular}[c]{@{}c@{}}BCIC IV 2a\\ (9 subjects,\\ 288 trials/subject)\end{tabular}} & FBCSP-SVM                                  & No                                   & 75.93 $\pm$ 14.93          & 74.49 $\pm$ 18.68                     & 58.09 $\pm$ 9.91            & 51.53 $\pm$ 24.01           \\
                                                                                                                     & Deep Convnet                               & Yes                                  & 63.72 $\pm$ 17.18          & 59.85 $\pm$ 22.17                     & 56.34 $\pm$ 8.86            & 30.62 $\pm$ 28.96           \\
                                                                                                                     & EEGNet-8,2                                 & Yes                                  & 65.93 $\pm$ 18.44          & 64.45 $\pm$ 26.23                     & 64.26 $\pm$ 11.03           & 60.19 $\pm$ 19.96           \\
                                                                                                                     & Spectral-Spatial CNN                       & No                                   & \textbf{76.91 $\pm$ 13.75} & \textbf{77.03 $\pm$ 15.41}            & \textbf{66.05 $\pm$ 13.70}  & \textbf{61.91 $\pm$ 20.31}  \\
                                                                                                                     & MIN2Net                                    & Yes                                  & 65.23 $\pm$ 16.14          & 64.72 $\pm$ 18.39                     & 60.03 $\pm$ 9.24            & 49.09 $\pm$ 23.28           \\ \midrule[0.1em]
            \multirow{5}{*}{\begin{tabular}[c]{@{}c@{}}SMR-BCI\\ (14 subjects,\\ 160 trials/subject)\end{tabular}}   & FBCSP-SVM                                  & No                                   & 74.50 $\pm$ 18.14          & \textbf{70.65 $\pm$ 23.64} & 62.64 $\pm$ 15.43           & 45.07 $\pm$ 34.93           \\
                                                                                                                     & Deep Convnet                               & Yes                                  & 61.40 $\pm$ 15.66          & 55.27 $\pm$ 22.00                     & 65.26 $\pm$ 16.83           & 54.38 $\pm$ 32.58           \\
                                                                                                                     & EEGNet-8,2                                 & Yes                                  & 67.76 $\pm$ 18.09          & 68.05 $\pm$ 21.11                     & 58.07 $\pm$ 11.45           & 34.43 $\pm$ 31.35           \\
                                                                                                                     & Spectral-Spatial CNN                       & No                                   & \textbf{76.76 $\pm$ 16.66} & 69.87 $\pm$ 28.15                     & \textbf{66.21 $\pm$ 15.15}  & 54.36 $\pm$ 31.21           \\
                                                                                                                     & MIN2Net                                    & Yes                                  & 65.90 $\pm$ 16.50          & 64.13 $\pm$ 17.66                     & 59.79 $\pm$ 13.72           & \textbf{61.10 $\pm$ 23.64}  \\ \midrule[0.1em]
            \multirow{5}{*}{\begin{tabular}[c]{@{}c@{}}OpenBMI\\ (54 subjects,\\ 400 trials/subject)\end{tabular}}   & FBCSP-SVM                                  & No                                   & \textbf{66.06 $\pm$ 16.58} & 64.66 $\pm$ 19.47                     & 64.96 $\pm$ 12.70           & 65.25 $\pm$ 15.14           \\
                                                                                                                     & Deep Convnet                               & Yes                                  & 60.31 $\pm$ 16.76          & 61.66 $\pm$ 18.17                     & 68.33 $\pm$ 15.33           & 70.20 $\pm$ 15.18           \\
                                                                                                                     & EEGNet-8,2                                 & Yes                                  & 60.41 $\pm$ 17.12          & 56.80 $\pm$ 23.54                     & 68.84 $\pm$ 13.87           & 70.39 $\pm$ 14.30           \\
                                                                                                                     & Spectral-Spatial CNN                       & No                                   & 65.19 $\pm$ 15.94          & \textbf{66.97 $\pm$ 16.71*}           & 68.27 $\pm$ 13.56           & 65.86 $\pm$ 17.37           \\
                                                                                                                     & MIN2Net                                    & Yes                                  & 61.03 $\pm$ 14.47          & 63.59 $\pm$ 14.52                     & \textbf{72.03 $\pm$ 14.04*} & \textbf{72.62 $\pm$ 14.14*} \\ \bottomrule[0.2em]
            \end{tabular}
    }
    \end{table*}

\begin{figure*}
\centering
\includegraphics[width=1.6\columnwidth]{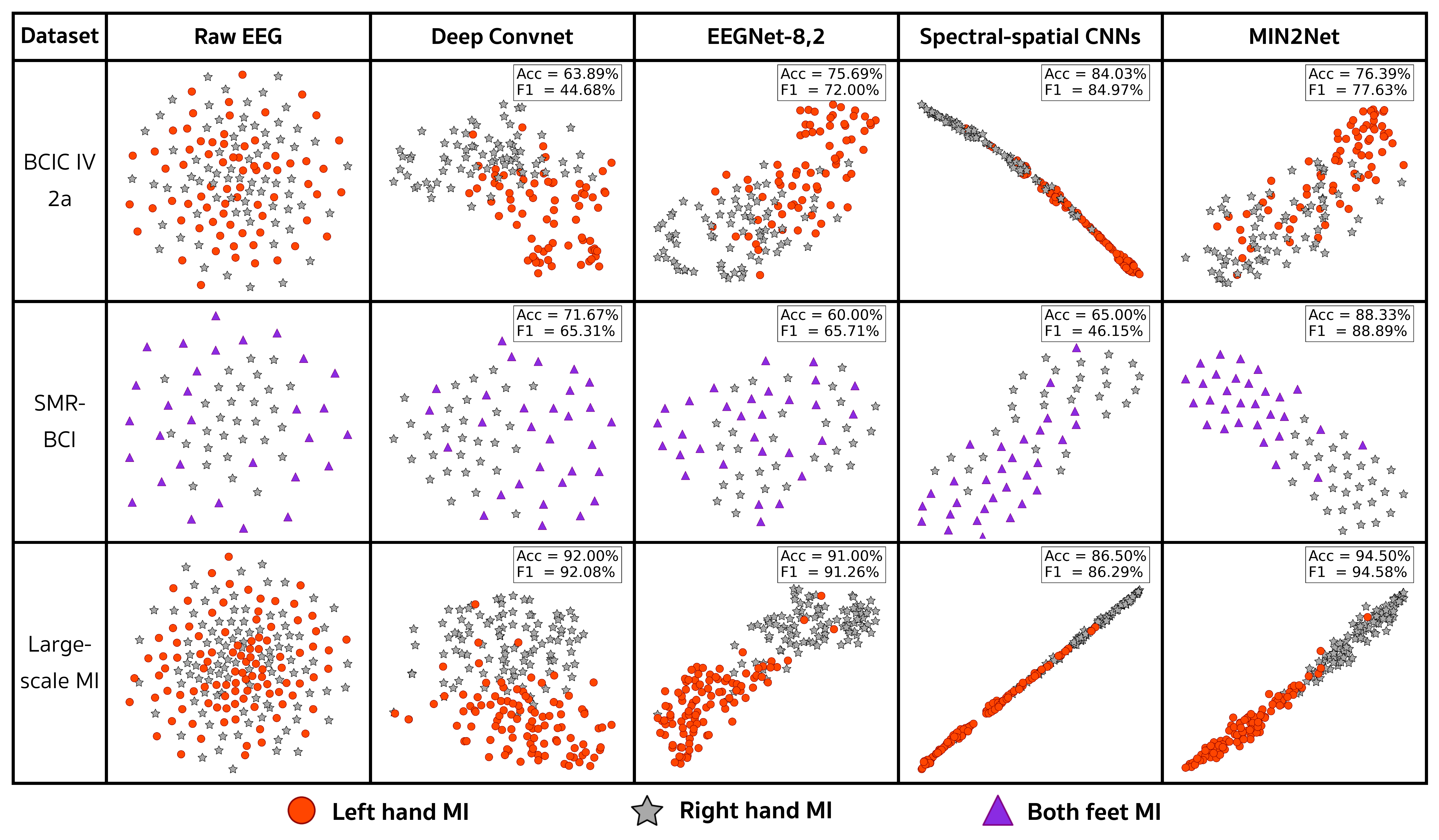}
\caption{Visualization of EEG features using two-dimensional $t$-SNE projection on the binary classification of EEG-MI data in the subject-independent manner. We picked one leaned EEG feature from one subject for each dataset.}
\label{fig:tsne_two_classes}
\end{figure*}

\subsection{Experiment II: Binary MI classification}
\autoref{tab:result_all} presents the overall performance of our MIN2Net and four baseline methods across all subjects for both subject-dependent and subject-independent settings. \textcolor{red}{It is observed that in the subject-independent manner, MIN2Net achieved the highest performance in terms of accuracy on the OpenBMI dataset and terms of F1-score on both SMR-BCI and OpenBMI datasets. Specifically, significant differences were seen among the accuracy and F1-score provided by MIN2Net and the other baseline methods in the OpenBMI dataset, $p < 0.05$. Considering the SMR-BCI dataset, the F1-score improvement of MIN2Net to the baseline methods was significant ($p < 0.05$) except for the Deep Convnet and Spectal-spatial CNN models. However, the overall performance of MIN2Net was lower than some baseline methods in the subject-independent over the BCIC IV 2a dataset. Furthermore, in the subject-dependent setting, the performance improvement of MIN2Net to the baseline methods was not found on three considered datasets.}  

\textcolor{red}{Moreover, we reveal the average training and prediction times for all subjects per epoch on all considered datasets as shown in \autoref{tab:result_time}. Note that the training time was identified as the duration time in each training iteration. Meanwhile, the prediction time was defined as the}
\textcolor{red}{duration time in each fold to classify all testing samples. The results also present the number of trainable parameters for MIN2Net and all baseline methods.}

\autoref{fig:tsne_two_classes} illustrates the $t$-SNE projection of the learned embedding features of all the datasets. The results display the 2-dimensional embedding features of MIN2Net and all baseline methods, considering all trials of one testing subject of each dataset. It can be seen that the embedding features from each class were more likely to be compactly clustered as well as attempted to preserve the relative distances among different clusters. 
 
\autoref{fig:train_samples} displays the variations of the classification F1-score with respect to the number of training samples. The number of training samples in all the datasets is represented on the $x$-axis, while the $y$-axis expresses the binary classification F1-score of MIN2Net and the two best baseline methods. It is demonstrated that increasing the number of training samples from 100 to 21200 samples can provide better classification F1-score, recommending as a significant factor to boost the final classification F1-score. Therefore, the classification F1-score of MIN2Net was shown to be higher in comparison to the F1-score of the two baseline methods when trained with a large number of training samples.   

\subsection{Experiment III: Multi-class MI classification}
\begin{figure}
\centering
\includegraphics[width=0.45\columnwidth]{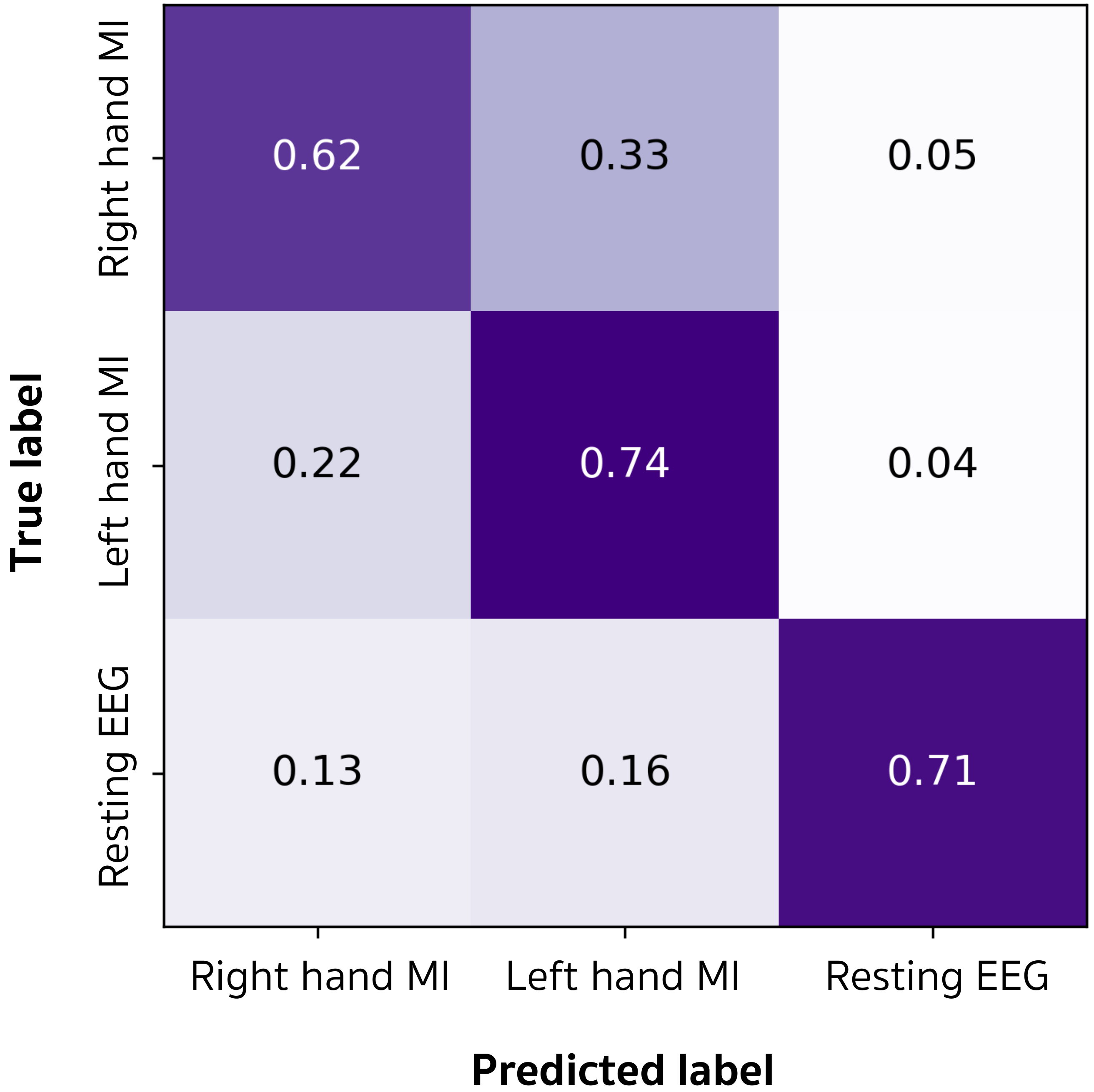}
\caption{Confusion matrix of three-class MI classification.}
\label{fig:confusion_three_classes}
\end{figure}

\autoref{tab:AE_triplet_results_3classes} exhibits the entire performance of the thee-class MI classification on OpenBMI dataset by comparing MIN2Net and all baseline approaches. \textcolor{red}{It was found that on the subject-independent manner, MIN2Net outperformed all baseline methods with the accuracy and F1-score of 68.81 $\pm$ 12.44\% and 68.04 $\pm$ 12.97\%, respectively. Moreover, there were significant differences in the accuracy and F1-score between MIN2Net and all baseline methods, $p < 0.05$.} \autoref{fig:confusion_three_classes} shows the confusion matrix of MIN2Net in the three-class MI classification on OpenBMI dataset. It was observed that in the subject-independent scenario, MIN2Net yielded the highest recall of resting EEG class and the lowest of right hand MI.

\autoref{fig:tsne_three_classes} reveals the scatter plots of the learned embedding features using $t$-SNE of the OpenBMI dataset. Results of our MIN2Net and the others are considered the three-class MI classification task from all trials of one representative subject. Similar to the $t$-SNE projection results in the binary classification, the learned embedding features from three different classes tended to be separated into three compact clusters and made an effort to maintain the relative distances among different clusters.

\begin{table*}
    \caption{\textcolor{red}{Time complexity (seconds per epoch) for all methods and number of trainable parameters for All deep learning approaches.}}
    \label{tab:result_time}
    \centering
    \resizebox{1.6\columnwidth}{!}{%
        \begin{tabular}{@{}ccccccc@{}}
        \toprule[0.2em]
        \multirow{2}{*}{\textbf{Datasets}} & \multirow{2}{*}{\textbf{Comparison Model}} & \multirow{2}{*}{\textit{\textbf{\#trainable params}}} & \multicolumn{2}{c}{\textbf{Subject-dependent}}                  & \multicolumn{2}{c}{\textbf{Subject-independent}}                \\ \cmidrule[0.1em](l){4-7} 
                                        &                                      &                                             & \textbf{Training Time} & \textbf{Prediction Time} & \textbf{Training Time} & \textbf{Prediction Time} \\ \midrule[0.1em]
    \multirow{5}{*}{BCIC IV 2a}        & FBCSP-SVM                              & -                                          & -                                & 0.0008                       & -                                & 0.0112                       \\
                                        & Deep Convnet                          & 151,027                                    & 0.1709                           & 0.1617                       & 0.2748                           & 0.1739                       \\
                                        & EEGNet-8,2                            & 5,162                                      & 0.1476                           & 0.1173                       & 0.3735                           & 0.0920                       \\
                                        & Spectral-Spatial CNN                  & 77,577,714                                 & 10.2031                          & 0.7600                       & 8.1334                           & 0.7444                       \\
                                        & MIN2Net                               & 55,232                                     & 0.2320                           & 0.1803                       & 0.4724                           & 0.2373                       \\ \midrule[0.1em]
     \multirow{5}{*}{SMR-BCI}           & FBCSP-SVM                             & -                                          & -                                & 0.0005                       & -                                & 0.0047                       \\
                                        & Deep Convnet                          & 150,302                                    & 0.1352                           & 0.1519                       & 0.2412                           & 0.1906                       \\
                                        & EEGNet-8,2                            & 5,082                                      & 0.1164                           & 0.1296                       & 0.3210                           & 0.1105                       \\
                                        & Spectral-Spatial CNN                  & 54,076,914                                 & 2.1321                           & 1.0257                       & 5.8785                           & 0.6688                       \\
                                        & MIN2Net                               & 38,297                                     & 0.1463                           & 0.2433                       & 0.4948                           & 0.2966                       \\ \midrule[0.1em]
     \multirow{5}{*}{OpenBMI}           & FBCSP-SVM                             & -                                          & -                                & 0.0020                       & -                                & 0.1906                       \\ 
                                        & Deep Convnet                          & 153,427                                    & 0.1804                           & 0.1618                       & 1.7497                           & 0.47345                       \\
                                        & EEGNet-8,2                            & 5,162                                      & 0.1882                           & 0.1439                       & 3.0951                           & 0.1372                       \\
                                        & Spectral-Spatial CNN                  & 77,577,714                                 & 2.2476                           & 1.0934                       & 11.9067                          & 0.8560                       \\
                                        & MIN2Net                               & 55,232                                     & 0.3527                           & 0.2851                       & 1.3626                           & 0.1043                       \\ \bottomrule[0.2em]
        \end{tabular}
    }
\end{table*}

\section{Discussion}

\begin{figure}
\centering
\includegraphics[width=0.9\columnwidth]{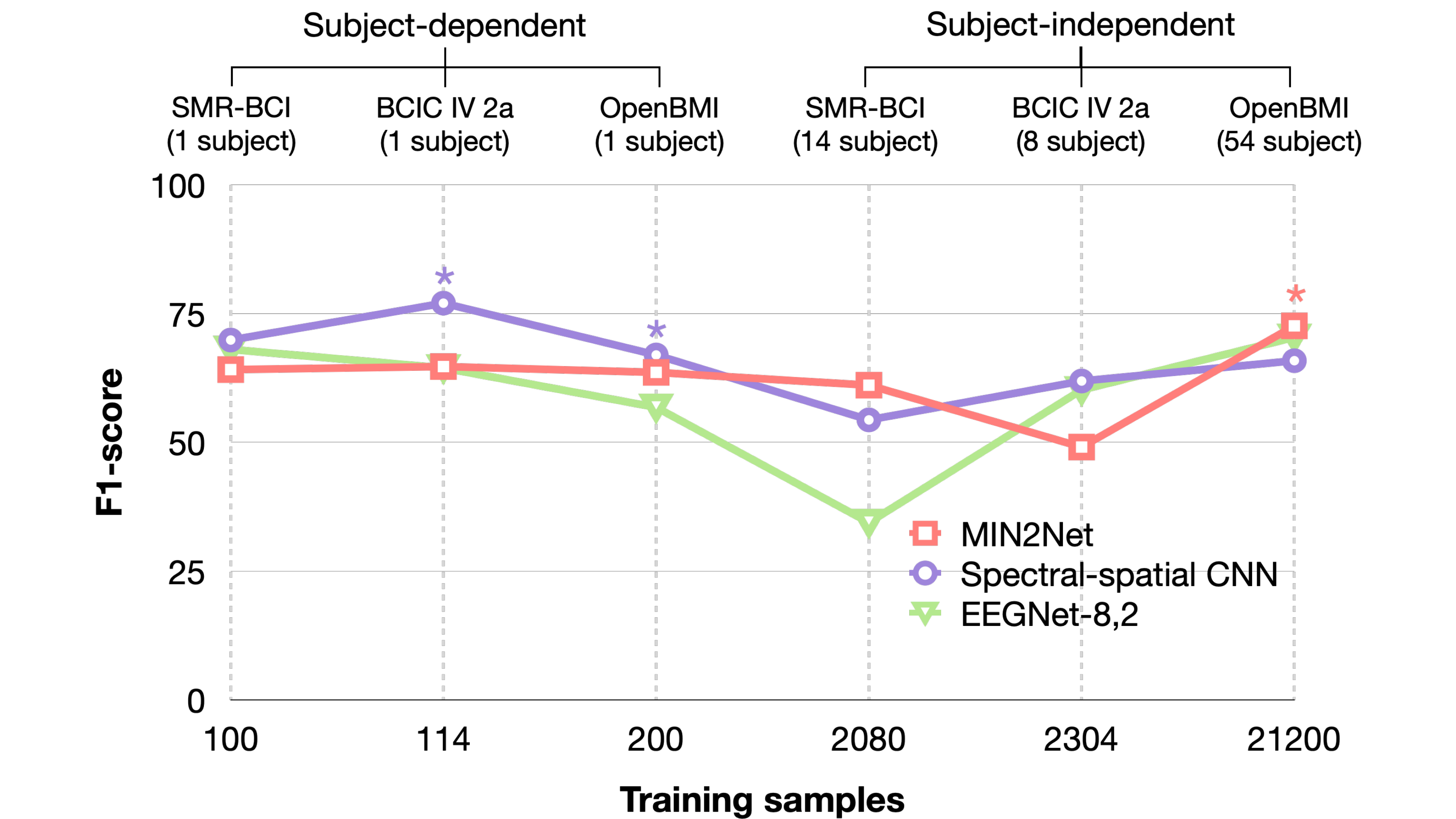}
\caption{Effect of the number of training samples on the binary classification performance across three considered methods.}
\label{fig:train_samples}

\end{figure}

\begin{table}
\caption{Classification accuracy and F1-score (in \%, $\pm$ SD)  on the OpenBMI dataset for the three-class classification of MI in the subject-independent manner. Bold denotes the best numerical values, and * represents the performance value which was significantly higher than all comparison pairs, $p < 0.05$.}
\label{tab:AE_triplet_results_3classes}
\centering
\resizebox{0.7\columnwidth}{!}{
    \begin{tabular}{@{}ccc@{}}
    \toprule[0.2em]
    \textbf{Algorithms}  & \textbf{Accuracy} & \textbf{F1-score} \\ \midrule[0.1em]
    FBCSP-SVM            & 50.71 $\pm$ 9.99        & 46.29 $\pm$ 12.40       \\ 
    Deep Convnet         & 54.04 $\pm$ 10.12	   & 49.77 $\pm$ 12.58      \\ 
    EEGNet-8,2        & 67.93 $\pm$ 11.94        & 66.41 $\pm$ 13.23       \\ 
    Spectral-spatial CNN & 64.67 $\pm$ 11.63        & 63.16 $\pm$ 12.53 \\ 
    MIN2Net      & \textbf{68.81 $\pm$ 12.44*}      & \textbf{68.04 $\pm$ 12.97*}     \\ \bottomrule[0.2em]
    \end{tabular}
}
\end{table}

\subsection{Effectiveness of Deep Metric Learning}

Deep learning (DL) has considerably contributed to the development of efficient MI-based BCI applications.
Among the DL techniques, end-to-end multi-task AE is
a powerful one to process raw EEG data due to the
combination of feature extraction and classification. However, AE can recognize instances rather than discriminating among classes. The current study investigated incorporating DML to a multi-task AE named MIN2Net. The role of DML is to solve the aforementioned limitation of AE. As the results in Table IV, DML using the triplet loss considerably affects the MIN2Net's performance, resulting in the classification results which are significantly higher than the MIN2Net–without triplet over two benchmark datasets. This suggests that DML can learn and improve discriminating of EEG data among different classes.

\subsection{Analysis of the Proposed Method}
Recently, the deep learning technique has attained popularity in BCI because it is capable of effectively learning the brain activity patterns from EEG data without using high complexity in EEG pre-processing\cite{8698218}. In this paper, our MIN2Net method's input is the time-domain EEG signals, filtered using a particular frequency band to eliminate high- and low-frequency artifacts. Based on Experiment I and II results, MIN2Net performs with higher accuracy than FBCSP and spectral-spatial CNN, which use multiple frequency bands to filter out artifacts. These results suggest that MIN2Net is robust to artifacts and offers higher classification results than the other baseline methods, utilizing simplistic EEG pre-processing only once. 

\begin{figure}
\centering
\includegraphics[width=0.8\columnwidth]{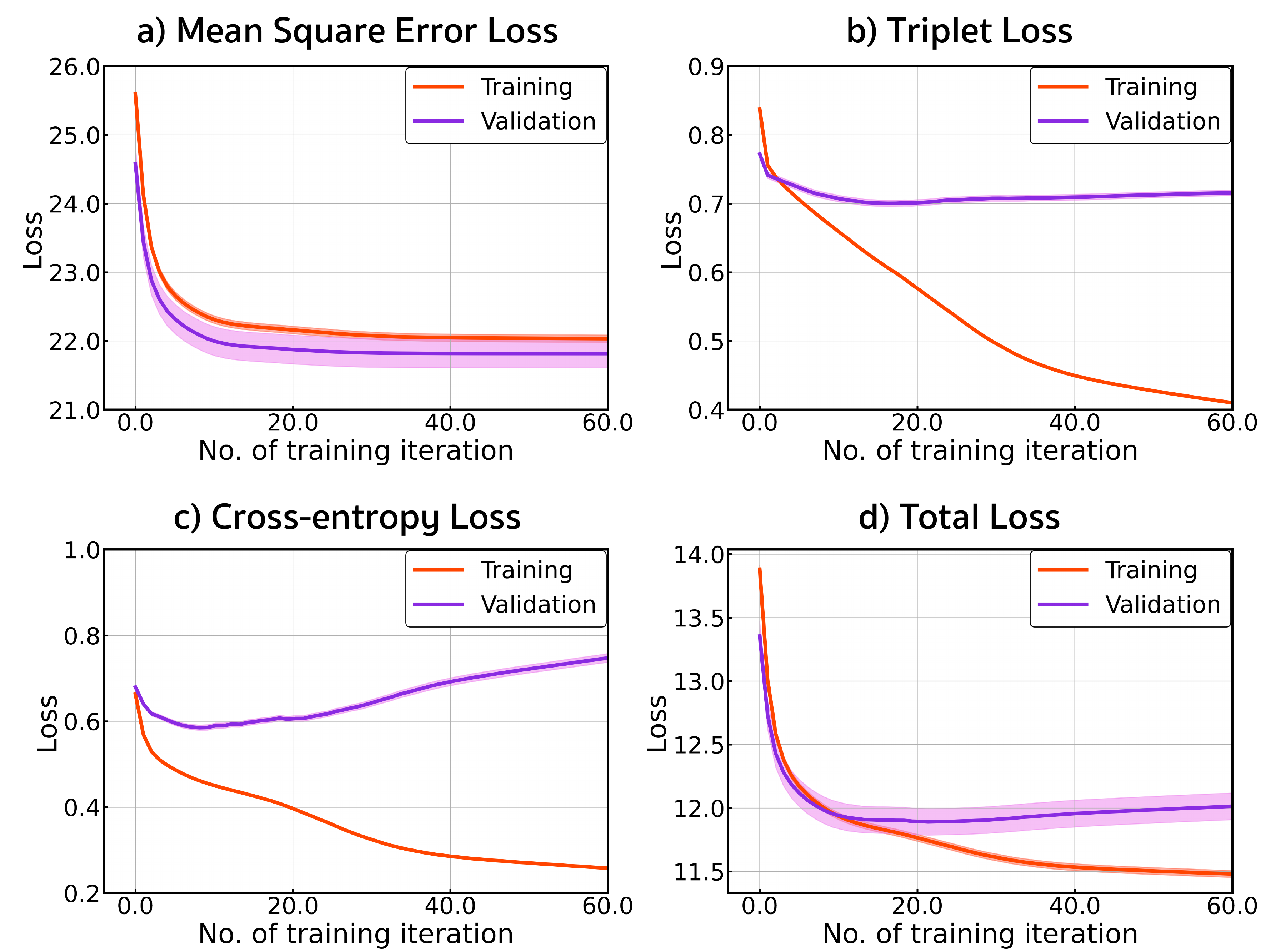}
\caption{Training and validation losses of our proposed MIN2Net on OpenBMI dataset. The plots show averaged losses with standard error of 54 subjects while d) total loss was weighed by 0.5, 0.5, 1.0 for a) mean square error  b) triplet and c) cross-entropy loss respectively.}
\label{fig:visualize_loss}
\end{figure}

To give insight into an internal perspective behind the optimization process of MIN2Net, we examined the changes of both training and validation losses in the binary classification during the training process of the OpenBMI dataset in a subject-independent manner. Four different losses of MIN2Net (MSE, triplet, cross-entropy, and the total losses) were monitored for 60 epochs from all subjects. It was observed that all the four losses converged around 15 epochs, as shown in \autoref{fig:visualize_loss}. Similar to the convergence process on the OpenBMI dataset, we also found signs of the convergence of these four losses within 60 epochs over the BCIC IV 2a and SMR-BCI datasets. \textcolor{red}{As the results in \autoref{tab:result_time}, MIN2Net has the 2\textsuperscript{nd} smallest size of trainable parameters on all considered datasets. Furthermore, the MIN2Net has a speed of training and prediction similar to the compact baseline models such as EEGNet-8,2 and Deep Convnet.}    

Regarding the module's weight ($\beta_1$, $\beta_2$ and $\beta_3$) for the loss function of MIN2Net shown in \autoref{tab:beta_param}, we can observe that the optimal performance obtains from different sets of the module's weight. The reason is related to the difference in the amount of data among all considered datasets, where the small dataset requires the small values in the module's weight to prevent overfitting. Meanwhile, all the module's weight values close to 1 are desirable for the large dataset to achieve optimal performance. We also found that when all three $\beta$ have the same value, there is a slight difference between the small and the large values on the large dataset. However, using the small dataset shows a considerable difference between the small and the large values as shown in the supplementary materials. 

\begin{figure}
\centering
\includegraphics[width=0.75\columnwidth]{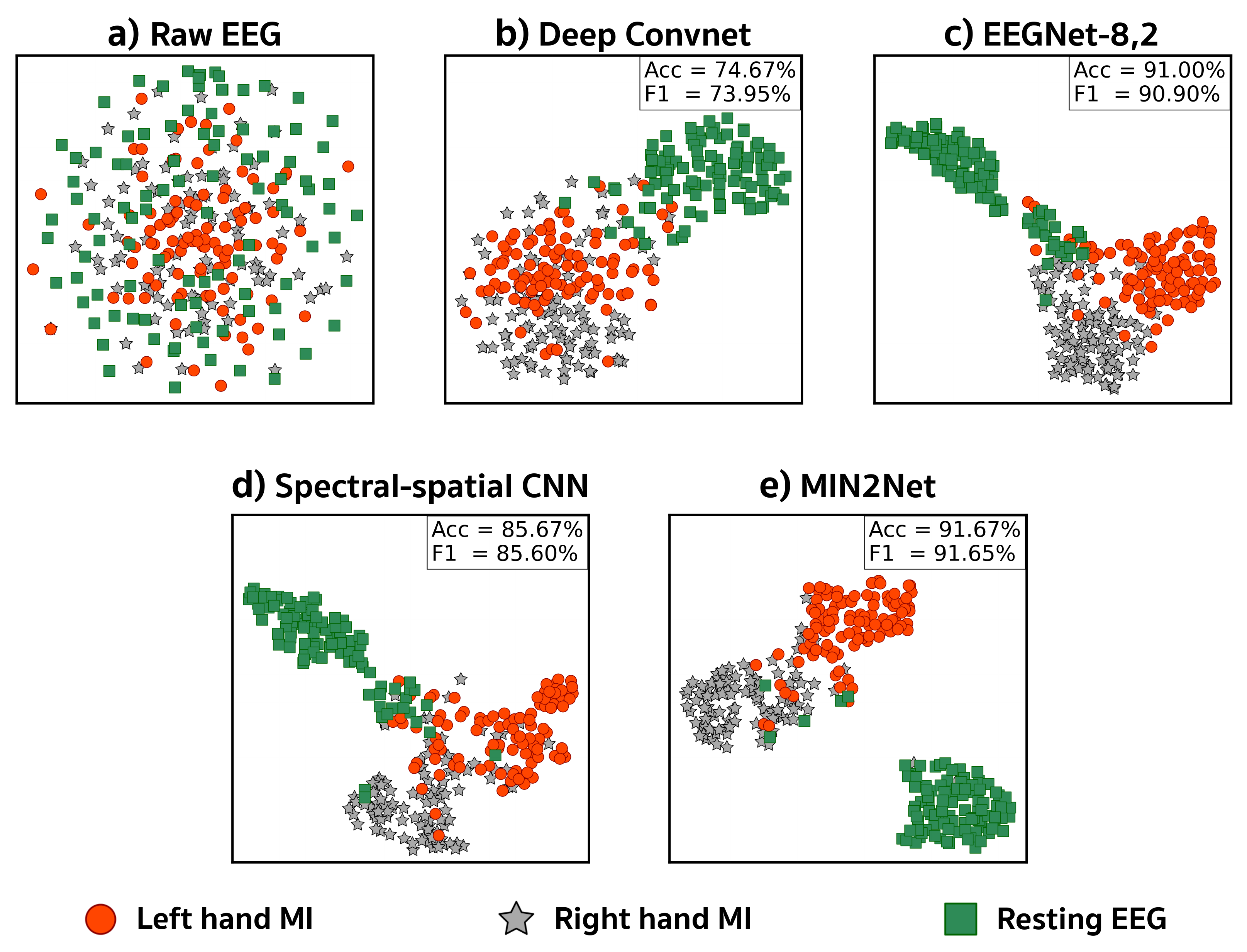}
\caption{Visualization of EEG features using two-dimensional $t$-SNE projection on the three-class classification of EEG-MI data in the subject-independent manner. We picked learned EEG features from one subject on OpenBMI dataset for visualization purpose.}
\label{fig:tsne_three_classes}
\end{figure}

\subsection{Analysis of the Comparison Performance}
The binary classification results on three benchmark datasets are listed in \autoref{tab:result_all}. \textcolor{red}{It can be observed that on SMR-BCI and OpenBMI datasets, MIN2Net outperforms all baseline methods in a subject-independent setting. Even though the SMR-BCI dataset's accuracy of MIN2Net is lower than some baseline methods, the F1-score of MIN2Net is higher than all baseline methods. According to the claims in \cite{f1_ref} that F1-score is more valuable than accuracy because it allows for both false positives and false negatives. Additionally, although the two benchmark datasets have different training samples, MIN2Net still results in the best performance in MI classification for both datasets. This investigation suggests that incorporating multi-task AE and DML, as done in MIN2Net, plays a vital role in extracting generalized EEG features for MI classification, resulting in excellent generalization performance on new subjects.} 


However, in the subject-dependent setting, the results reveal suboptimal classification performance of MIN2Net on three benchmark datasets. The reason is that MIN2Net has a small number of training samples where the few training samples are from only one subject, so that this leads to an overfitting problem. Generally, the deep learning approach requires a large amount of data to efficiently train a generalized model for preventing an overfitting problem\cite{9177281}. As illustrated in \autoref{fig:train_samples}, the result confirmed that MIN2Net performed with higher accuracy than the others when using a large number of training samples. \textcolor{red}{Therefore, one suggestive solution to alleviate this overfitting issue is to increase training samples in the subject-dependent setting by data augmentation methods. We believed that more training samples could help the model capture generalized features and improve the MI classification performance. To prove this, we experimented with how the subject-dependent setting with data augmentation enhanced the MIN2Net performance on all considered datasets. When we applied the data augmentation methods for the EEG data in the subject-dependent setting, it was found that significantly higher classification performance, as shown in \autoref{tab:augmentation} in Appendix.}


\subsection{Visualization of the Learned Latent Representation}
\autoref{fig:tsne_two_classes} shows the $t$-SNE visualization results from the binary MI classification in the subject-independent setting.
The 2D embedding features from raw EEG data in each dataset demonstrates the non-relative features among MI classes, as depicted in \autoref{fig:tsne_two_classes} (2\textsuperscript{nd} column). \autoref{fig:tsne_two_classes} (3\textsuperscript{rd} column), (4\textsuperscript{th} column), and (5\textsuperscript{th} column) present the 2D embedding features of high dimensional embedding features at the input of the final FC layer for the baseline DeepConvNet, EEGNet-8,2, and Spectral-spatial CNN methods, indicating that their embedding features of MI signals from different classes were more likely to overlap with each other. It can be seen that the latent features produced by our MIN2Net method, as depicted in \autoref{fig:tsne_two_classes} (6\textsuperscript{th} column), reveal highly discriminative patterns over SMR-BCI and OpenBMI datasets. Similarly, in the multi-class classification, the latent features extracted by MIN2Net also provide predominantly discriminative patterns on the OpenBMI dataset compared to the other baseline methods, as shown in \autoref{fig:tsne_three_classes}. Thus, this provides evidence that our MIN2Net method produces superior performance owing to the greater quality representation of the MI in the learned latent features.  

\subsection{Feasibility in online BCI systems}
\textcolor{red}{To further evaluate the feasibility and pseudo-online performance of MIN2Net, we established an experiment on the OpenBMI dataset, exhibiting the classification among resting EEG, left hand MI, and right hand MI. The results demonstrate that MIN2Net significantly outperforms all baseline methods in the subject-independent setting, as depicted in \autoref{tab:AE_triplet_results_3classes}. Additionally, MIN2Net yields an acceptable misclassification rate in the classification of three-class MI, as depicted in \autoref{fig:confusion_three_classes}. The present results confirm the possibility of using the MIN2Net in two promising aspects. First, MIN2Net has the capability of classifying multi-class MI-EEG. Second, MIN2Net could integrate with online BCI applications, providing various movement intentions for users, such as stop moving, left-hand grasping, or right-hand grasping.}

\textcolor{red}{Moreover, these research findings provide foundations in developing BCI applications based on a calibration-free method. The proposed method can be used as a pre-trained model, where the model is pre-trained before being applied by a new user. The latency of the proposed method is considered as a prediction time in the testing session. With the subject-independent setting, the BCI application considers the prediction time rather than the training time. According to \autoref{tab:result_time}, it is found that the prediction time to classify all testing trials is 0.2373 s, 0.2966 s, and 0.1043 s for BCIC IV 2a, SMR-BCI, and OpenBMI datasets, respectively. As shown in \autoref{fig:train_samples}, when few subjects are used for training, we could observe that the proposed model provides suboptimal performance. On the other hand, training with more subjects could improve the overall performance of the proposed method. Therefore, once we perceive a new user as an outstanding BCI user, we can include that user in our dataset to retrain the proposed model to achieve better classification performance.}
\\

\subsection{Future Direction}
Although MIN2Net achieves a promising classification result, there are still several rooms for further improvement. Firstly, the DML module is based on triplet loss, and there are numerous new loss functions developed for DML that are more attractive to investigate\cite{contrastive_paper,quadruplet_paper}. Thus, in the future study, we will focus on incorporating these new loss functions into multi-task AE to further improve classification performance. Secondly, we can explore the utilization of MIN2Net on other EEG measurements, such as SSVEP, MRCPs, and ERP. MIN2Nets might be helpful in the extraction of the most discriminative features for classification. Finally, a transfer learning framework based on a fast adaptation procedure will be considered in our future work to thoroughly investigate the possibility of our MIN2Net\cite{ZHANG20211, 7802578}. 

\section{Conclusion}
This study proposed MIN2Net, a novel end-to-end multi-task learning, for classifying motor imagery EEG signals. MIN2Net is developed by integrating an autoencoder, a deep metric learning, and a supervised classifier, which learns to compress, discriminate embedded EEG, and classify EEG simultaneously. We compared the binary classification performance of MIN2Net with four different deep learning algorithms on three benchmark datasets. The classification results revealed that MIN2Net significantly outperformed the developed baselines in the subject-independent experiments on SMR-BCI and OpenBMI datasets. Moreover, we obtained promising experimental results from three-class EEG-MI classification (left hand MI vs. right hand MI vs. resting EEG). This indicates the possibility and practicality of using this model toward developing real-world applications.

\begin{appendices}
\section*{appendix}
\textcolor{red}{As the results from \autoref{tab:result_all}, MIN2Net provided suboptimal classification performance in the subject-independent manners owing to training with few training samples from one subject. To overcome this issue, data augmentation methods were adopted to increase the training samples in the subject-independent manners. Jittering (Jitter), Magnitude-Warping (MagW), Scaling (Scale), Time-Warping (TimeW) and Permutation (Perm), which were introduced by \cite{data_augment} for bio-signals, were applied to create new ones. The results from using data augmentation methods are illustrated in \autoref{tab:augmentation}.}
\begin{table}[t]
\caption{Classification accuracy and F1-score (in \%, $\pm$ SD) of MIN2Net using different subject-dependent manners. Bold denotes the best numerical values, and * represents the performance value which was significantly higher than all comparison pairs, $p < 0.05$.}
\label{tab:augmentation}
\centering
\resizebox{0.9\columnwidth}{!}{
    \begin{tabular}{@{}cccc@{}}
    \toprule[0.2em]
    \textbf{Dataset}            & \textbf{Comparison Manner}                             & \textbf{Accuracy}                     & \textbf{F1-score}                     \\ \midrule[0.1em]
    \multirow{2}{*}{BCIC IV 2a} & without aug.                                      & 65.23 $\pm$ 16.14                     & 64.72 $\pm$ 18.39                     \\ 
                                & with aug. & \multicolumn{1}{l}{\textbf{70.09 $\pm$ 16.87*}} & \multicolumn{1}{l}{\textbf{70.44 $\pm$ 16.49*}} \\ \midrule[0.1em]
    \multirow{2}{*}{SMR-BCI}    & without aug.                                       & 65.90 $\pm$ 16.50                     & 64.13 $\pm$ 17.66                     \\ 
                                & with aug. & \multicolumn{1}{l}{\textbf{72.95 $\pm$ 15.76*}} & \multicolumn{1}{l}{\textbf{69.51 $\pm$ 20.00*}} \\ \midrule[0.1em]
    \multirow{2}{*}{OpenBMI}    & without aug.                                       & 61.03 $\pm$ 14.47                     & 63.59 $\pm$ 14.52                     \\ 
                                & with aug. & \multicolumn{1}{l}{\textbf{66.51 $\pm$ 15.53*}} & \multicolumn{1}{l}{\textbf{68.47 $\pm$ 14.65*}} \\ \bottomrule[0.2em]
    \end{tabular}
}
\end{table}
\end{appendices}
\bibliography{References}

\begin{thebibliography}{10}
\providecommand{\url}[1]{#1}
\csname url@samestyle\endcsname
\providecommand{\newblock}{\relax}
\providecommand{\bibinfo}[2]{#2}
\providecommand{\BIBentrySTDinterwordspacing}{\spaceskip=0pt\relax}
\providecommand{\BIBentryALTinterwordstretchfactor}{4}
\providecommand{\BIBentryALTinterwordspacing}{\spaceskip=\fontdimen2\font plus
\BIBentryALTinterwordstretchfactor\fontdimen3\font minus
  \fontdimen4\font\relax}
\providecommand{\BIBforeignlanguage}[2]{{%
\expandafter\ifx\csname l@#1\endcsname\relax
\typeout{** WARNING: IEEEtran.bst: No hyphenation pattern has been}%
\typeout{** loaded for the language `#1'. Using the pattern for}%
\typeout{** the default language instead.}%
\else
\language=\csname l@#1\endcsname
\fi
#2}}
\providecommand{\BIBdecl}{\relax}
\BIBdecl

\bibitem{MCFARLAND2017194}
D.~{McFarland} and J.~{Wolpaw}, ``Eeg-based brain–computer interfaces,''
  \emph{Current Opinion in Biomedical Engineering}, vol.~4, pp. 194--200, 2017.

\bibitem{8945233}
P.~{Sawangjai}, S.~{Hompoonsup}, P.~{Leelaarporn}, S.~{Kongwudhikunakorn}, and
  T.~{Wilaiprasitporn}, ``Consumer grade eeg measuring sensors as research
  tools: A review,'' \emph{IEEE Sensors Journal}, vol.~20, no.~8, pp.
  3996--4024, 2020.

\bibitem{6737255}
H.~{Cecotti}, M.~P. {Eckstein}, and B.~{Giesbrecht}, ``Single-trial
  classification of event-related potentials in rapid serial visual
  presentation tasks using supervised spatial filtering,'' \emph{IEEE
  Transactions on Neural Networks and Learning Systems}, vol.~25, no.~11, pp.
  2030--2042, 2014.

\bibitem{6683068}
R.~{Chai}, S.~H. {Ling}, G.~P. {Hunter}, Y.~{Tran}, and H.~T. {Nguyen},
  ``Brain–computer interface classifier for wheelchair commands using neural
  network with fuzzy particle swarm optimization,'' \emph{IEEE Journal of
  Biomedical and Health Informatics}, vol.~18, no.~5, pp. 1614--1624, 2014.

\bibitem{8926475}
P.~{Autthasan}, X.~{Du}, J.~{Arnin}, S.~{Lamyai}, M.~{Perera},
  S.~{Itthipuripat}, T.~{Yagi}, P.~{Manoonpong}, and T.~{Wilaiprasitporn}, ``A
  single-channel consumer-grade eeg device for brain–computer interface:
  Enhancing detection of ssvep and its amplitude modulation,'' \emph{IEEE
  Sensors Journal}, vol.~20, no.~6, pp. 3366--3378, 2020.

\bibitem{6955799}
Y.~{Zou}, V.~{Nathan}, and R.~{Jafari}, ``Automatic identification of
  artifact-related independent components for artifact removal in eeg
  recordings,'' \emph{IEEE Journal of Biomedical and Health Informatics},
  vol.~20, no.~1, pp. 73--81, 2016.

\bibitem{8960436}
J.~{Jeong}, N.~{Kwak}, C.~{Guan}, and S.~{Lee}, ``Decoding movement-related
  cortical potentials based on subject-dependent and section-wise spectral
  filtering,'' \emph{IEEE Transactions on Neural Systems and Rehabilitation
  Engineering}, vol.~28, no.~3, pp. 687--698, 2020.

\bibitem{9130151}
R.~{Chaisaen}, P.~{Autthasan}, N.~{Mingchinda}, P.~{Leelaarporn},
  N.~{Kunaseth}, S.~{Tammajarung}, P.~{Manoonpong}, S.~C. {Mukhopadhyay}, and
  T.~{Wilaiprasitporn}, ``Decoding eeg rhythms during action observation, motor
  imagery, and execution for standing and sitting,'' \emph{IEEE Sensors
  Journal}, vol.~20, no.~22, pp. 13\,776--13\,786, 2020.

\bibitem{PFURTSCHELLER19991842}
G.~Pfurtscheller and F.~{Lopes da Silva}, ``Event-related eeg/meg
  synchronization and desynchronization: basic principles,'' \emph{Clinical
  Neurophysiology}, vol. 110, no.~11, pp. 1842 -- 1857, 1999.

\bibitem{7802578}
K.~K. {Ang} and C.~{Guan}, ``Eeg-based strategies to detect motor imagery for
  control and rehabilitation,'' \emph{IEEE Transactions on Neural Systems and
  Rehabilitation Engineering}, vol.~25, no.~4, pp. 392--401, 2017.

\bibitem{8737742}
A.~M. {Azab}, L.~{Mihaylova}, K.~K. {Ang}, and M.~{Arvaneh}, ``Weighted
  transfer learning for improving motor imagery-based brain–computer
  interface,'' \emph{IEEE Transactions on Neural Systems and Rehabilitation
  Engineering}, vol.~27, no.~7, pp. 1352--1359, 2019.

\bibitem{ZHANG20211}
K.~Zhang, N.~Robinson, S.-W. Lee, and C.~Guan, ``Adaptive transfer learning for
  eeg motor imagery classification with deep convolutional neural network,''
  \emph{Neural Networks}, vol. 136, pp. 1 -- 10, 2021.

\bibitem{Hinton504}
G.~E. Hinton and R.~R. Salakhutdinov, ``Reducing the dimensionality of data
  with neural networks,'' \emph{Science}, vol. 313, no. 5786, pp. 504--507,
  2006.

\bibitem{8897723}
O.~Y. {Kwon}, M.~H. {Lee}, C.~{Guan}, and S.~W. {Lee}, ``Subject-independent
  brain–computer interfaces based on deep convolutional neural networks,''
  \emph{IEEE Transactions on Neural Networks and Learning Systems}, vol.~31,
  no.~10, pp. 3839--3852, 2020.

\bibitem{1506823}
D.~J. {McFarland} and J.~R. {Wolpaw}, ``Sensorimotor rhythm-based
  brain-computer interface (bci): feature selection by regression improves
  performance,'' \emph{IEEE Transactions on Neural Systems and Rehabilitation
  Engineering}, vol.~13, no.~3, pp. 372--379, 2005.

\bibitem{939829}
G.~{Pfurtscheller} and C.~{Neuper}, ``Motor imagery and direct brain-computer
  communication,'' \emph{Proceedings of the IEEE}, vol.~89, no.~7, pp.
  1123--1134, 2001.

\bibitem{JIAO2018582}
Z.~Jiao, X.~Gao, Y.~Wang, J.~Li, and H.~Xu, ``Deep convolutional neural
  networks for mental load classification based on eeg data,'' \emph{Pattern
  Recognition}, vol.~76, pp. 582 -- 595, 2018.

\bibitem{facenet}
F.~{Schroff}, D.~{Kalenichenko}, and J.~{Philbin}, ``Facenet: A unified
  embedding for face recognition and clustering,'' in \emph{2015 IEEE
  Conference on Computer Vision and Pattern Recognition (CVPR)}, 2015, pp.
  815--823.

\bibitem{Wang_2019_CVPR}
X.~Wang, X.~Han, W.~Huang, D.~Dong, and M.~R. Scott, ``Multi-similarity loss
  with general pair weighting for deep metric learning,'' in \emph{Proceedings
  of the IEEE/CVF Conference on Computer Vision and Pattern Recognition
  (CVPR)}, June 2019.

\bibitem{tonio_paper}
R.~T. Schirrmeister, J.~T. Springenberg, L.~D.~J. Fiederer, M.~Glasstetter,
  K.~Eggensperger, M.~Tangermann, F.~Hutter, W.~Burgard, and T.~Ball, ``Deep
  learning with convolutional neural networks for eeg decoding and
  visualization,'' \emph{Human Brain Mapping}, vol.~38, no.~11, pp. 5391--5420,
  2017.

\bibitem{8310961}
S.~{Sakhavi}, C.~{Guan}, and S.~{Yan}, ``Learning temporal information for
  brain-computer interface using convolutional neural networks,'' \emph{IEEE
  Transactions on Neural Networks and Learning Systems}, vol.~29, no.~11, pp.
  5619--5629, 2018.

\bibitem{Lawhern_2018}
V.~J. Lawhern, A.~J. Solon, N.~R. Waytowich, S.~M. Gordon, C.~P. Hung, and
  B.~J. Lance, ``{EEGNet}: a compact convolutional neural network for
  {EEG}-based brain{\textendash}computer interfaces,'' \emph{Journal of Neural
  Engineering}, vol.~15, no.~5, p. 056013, jul 2018.

\bibitem{8745473}
T.~{Wilaiprasitporn}, A.~{Ditthapron}, K.~{Matchaparn}, T.~{Tongbuasirilai},
  N.~{Banluesombatkul}, and E.~{Chuangsuwanich}, ``Affective eeg-based person
  identification using the deep learning approach,'' \emph{IEEE Transactions on
  Cognitive and Developmental Systems}, vol.~12, no.~3, pp. 486--496, 2020.

\bibitem{8556024}
P.~{Zhang}, X.~{Wang}, W.~{Zhang}, and J.~{Chen}, ``Learning
  spatial–spectral–temporal eeg features with recurrent 3d convolutional
  neural networks for cross-task mental workload assessment,'' \emph{IEEE
  Transactions on Neural Systems and Rehabilitation Engineering}, vol.~27,
  no.~1, pp. 31--42, 2019.

\bibitem{7752836}
A.~{Gogna}, A.~{Majumdar}, and R.~{Ward}, ``Semi-supervised stacked label
  consistent autoencoder for reconstruction and analysis of biomedical
  signals,'' \emph{IEEE Transactions on Biomedical Engineering}, vol.~64,
  no.~9, pp. 2196--2205, 2017.

\bibitem{8723080}
A.~{Ditthapron}, N.~{Banluesombatkul}, S.~{Ketrat}, E.~{Chuangsuwanich}, and
  T.~{Wilaiprasitporn}, ``Universal joint feature extraction for p300 eeg
  classification using multi-task autoencoder,'' \emph{IEEE Access}, vol.~7,
  pp. 68\,415--68\,428, 2019.

\bibitem{895946}
H.~{Ramoser}, J.~{Muller-Gerking}, and G.~{Pfurtscheller}, ``Optimal spatial
  filtering of single trial eeg during imagined hand movement,'' \emph{IEEE
  Transactions on Rehabilitation Engineering}, vol.~8, no.~4, pp. 441--446,
  2000.

\bibitem{4634130}
{Kai Keng Ang}, {Zheng Yang Chin}, {Haihong Zhang}, and {Cuntai Guan}, ``Filter
  bank common spatial pattern (fbcsp) in brain-computer interface,'' in
  \emph{2008 IEEE International Joint Conference on Neural Networks (IEEE World
  Congress on Computational Intelligence)}, 2008, pp. 2390--2397.

\bibitem{second_fbcsp}
K.~K. Ang, Z.~Y. Chin, C.~Wang, C.~Guan, and H.~Zhang, ``Filter bank common
  spatial pattern algorithm on bci competition iv datasets 2a and 2b,''
  \emph{Frontiers in Neuroscience}, vol.~6, p.~39, 2012.

\bibitem{5593210}
F.~{Lotte} and C.~{Guan}, ``Regularizing common spatial patterns to improve bci
  designs: Unified theory and new algorithms,'' \emph{IEEE Transactions on
  Biomedical Engineering}, vol.~58, no.~2, pp. 355--362, 2011.

\bibitem{9272754}
Y.~Jiao, T.~Zhou, L.~Yao, G.~Zhou, X.~Wang, and Y.~Zhang, ``Multi-view
  multi-scale optimization of feature representation for eeg classification
  improvement,'' \emph{IEEE Transactions on Neural Systems and Rehabilitation
  Engineering}, vol.~28, no.~12, pp. 2589--2597, 2020.

\bibitem{9349966}
Y.~Zhang, T.~Zhou, W.~Wu, H.~Xie, H.~Zhu, G.~Zhou, and A.~Cichocki, ``Improving
  eeg decoding via clustering-based multitask feature learning,'' \emph{IEEE
  Transactions on Neural Networks and Learning Systems}, pp. 1--11, 2021.

\bibitem{NECO_a_00592}
A.~Llera, V.~Gómez, and H.~J. Kappen, ``{Adaptive Multiclass Classification
  for Brain Computer Interfaces},'' \emph{Neural Computation}, vol.~26, no.~6,
  pp. 1108--1127, 06 2014.

\bibitem{8353425}
Y.~Jiao, Y.~Zhang, X.~Chen, E.~Yin, J.~Jin, X.~Wang, and A.~Cichocki, ``Sparse
  group representation model for motor imagery eeg classification,'' \emph{IEEE
  Journal of Biomedical and Health Informatics}, vol.~23, no.~2, pp. 631--641,
  2019.

\bibitem{9175874}
R.~{Mane}, N.~{Robinson}, A.~P. {Vinod}, S.~W. {Lee}, and C.~{Guan}, ``A
  multi-view cnn with novel variance layer for motor imagery brain computer
  interface,'' in \emph{2020 42nd Annual International Conference of the IEEE
  Engineering in Medicine Biology Society (EMBC)}, 2020, pp. 2950--2953.

\bibitem{lee_dataset}
M.-H. Lee, O.-Y. Kwon, Y.-J. Kim, H.-K. Kim, Y.-E. Lee, J.~Williamson,
  S.~Fazli, and S.-W. Lee, ``{EEG dataset and OpenBMI toolbox for three BCI
  paradigms: an investigation into BCI illiteracy},'' \emph{GigaScience},
  vol.~8, no.~5, 01 2019.

\bibitem{Ge_2018_ECCV}
W.~Ge, ``Deep metric learning with hierarchical triplet loss,'' in
  \emph{Proceedings of the European Conference on Computer Vision (ECCV)},
  September 2018.

\bibitem{contrastive_paper}
K.~Prannay, T.~Piotr, W.~Chen, S.~Aaron, T.~Yonglong, I.~Phillip, M.~Aaron,
  L.~Ce, and K.~Dilip, ``Supervised contrastive learning,'' \emph{NeurIPS
  2020}, 2020.

\bibitem{7298682}
F.~{Schroff}, D.~{Kalenichenko}, and J.~{Philbin}, ``Facenet: A unified
  embedding for face recognition and clustering,'' in \emph{2015 IEEE
  Conference on Computer Vision and Pattern Recognition (CVPR)}, 2015, pp.
  815--823.

\bibitem{quadruplet_paper}
W.~Chen, X.~Chen, J.~Zhang, and K.~Huang, ``\BIBforeignlanguage{English}{Beyond
  triplet loss: a deep quadruplet network for person re-identification},'' in
  \emph{\BIBforeignlanguage{English}{Proceedings of IEEE International
  Conference on Computer Vision and Pattern Recognition}}.\hskip 1em plus 0.5em
  minus 0.4em\relax IEEE, 2017, pp. 1320--1329.

\bibitem{8257016}
R.~{Thiyagarajan}, C.~{Curro}, and S.~{Keene}, ``A learned embedding space for
  eeg signal clustering,'' in \emph{2017 IEEE Signal Processing in Medicine and
  Biology Symposium (SPMB)}, 2017, pp. 1--4.

\bibitem{9116935}
H.~{Alwasiti}, M.~Z. {Yusoff}, and K.~{Raza}, ``Motor imagery classification
  for brain computer interface using deep metric learning,'' \emph{IEEE
  Access}, vol.~8, pp. 109\,949--109\,963, 2020.

\bibitem{fisrt_ae}
D.~E. Rumelhart, G.~E. Hinton, and R.~J. Williams, \emph{Learning Internal
  Representations by Error Propagation}.\hskip 1em plus 0.5em minus 0.4em\relax
  Cambridge, MA, USA: MIT Press, 1986, p. 318–362.

\bibitem{552239}
G.~{Antoniol} and P.~{Tonella}, ``Eeg data compression techniques,'' \emph{IEEE
  Transactions on Biomedical Engineering}, vol.~44, no.~2, pp. 105--114, 1997.

\bibitem{SAE_paper}
P.~Vincent, H.~Larochelle, I.~Lajoie, Y.~Bengio, and P.-A. Manzagol, ``Stacked
  denoising autoencoders: Learning useful representations in a deep network
  with a local denoising criterion,'' \emph{J. Mach. Learn. Res.}, vol.~11, p.
  3371–3408, Dec. 2010.

\bibitem{8429921}
Y.~{Qiu}, W.~{Zhou}, N.~{Yu}, and P.~{Du}, ``Denoising sparse autoencoder-based
  ictal eeg classification,'' \emph{IEEE Transactions on Neural Systems and
  Rehabilitation Engineering}, vol.~26, no.~9, pp. 1717--1726, 2018.

\bibitem{8416796}
M.~{Wang}, S.~{Abdelfattah}, N.~{Moustafa}, and J.~{Hu}, ``Deep gaussian
  mixture-hidden markov model for classification of eeg signals,'' \emph{IEEE
  Transactions on Emerging Topics in Computational Intelligence}, vol.~2,
  no.~4, pp. 278--287, 2018.

\bibitem{BCIC_IV_2a}
M.~Tangermann, K.-R. Müller, A.~Aertsen, N.~Birbaumer, C.~Braun, C.~Brunner,
  R.~Leeb, C.~Mehring, K.~Miller, G.~Mueller-Putz, G.~Nolte, G.~Pfurtscheller,
  H.~Preissl, G.~Schalk, A.~Schlögl, C.~Vidaurre, S.~Waldert, and
  B.~Blankertz, ``Review of the bci competition iv,'' \emph{Frontiers in
  Neuroscience}, vol.~6, p.~55, 2012.

\bibitem{BCIC_SMR}
D.~Steyrl, R.~Scherer, O.~F{\"o}rstner, and G.~M{\"u}ller-Putz,
  ``\BIBforeignlanguage{English}{Motor imagery brain-computer interfaces:
  Random forests vs regularized lda - non-linear beats linear},'' in
  \emph{\BIBforeignlanguage{English}{Proceedings of the 6th International
  Brain-Computer Interface Conference Graz 2014}}, 2014, pp. 061--1--061--4.

\bibitem{Schroff_2015}
F.~Schroff, D.~Kalenichenko, and J.~Philbin, ``Facenet: A unified embedding for
  face recognition and clustering,'' \emph{2015 IEEE Conference on Computer
  Vision and Pattern Recognition (CVPR)}, Jun 2015.

\bibitem{mne_package}
A.~Gramfort, M.~Luessi, E.~Larson, D.~Engemann, D.~Strohmeier, C.~Brodbeck,
  R.~Goj, M.~Jas, T.~Brooks, L.~Parkkonen, and M.~Hämäläinen, ``Meg and eeg
  data analysis with mne-python,'' \emph{Frontiers in Neuroscience}, vol.~7, p.
  267, 2013.

\bibitem{vandermaaten08a}
L.~van~der Maaten and G.~Hinton, ``Visualizing data using t-sne,''
  \emph{Journal of Machine Learning Research}, vol.~9, no.~86, pp. 2579--2605,
  2008.

\bibitem{8698218}
D.~{Zhang}, L.~{Yao}, K.~{Chen}, S.~{Wang}, X.~{Chang}, and Y.~{Liu}, ``Making
  sense of spatio-temporal preserving representations for eeg-based human
  intention recognition,'' \emph{IEEE Transactions on Cybernetics}, vol.~50,
  no.~7, pp. 3033--3044, 2020.

\bibitem{f1_ref}
S.~Bhattacharyya, A.~Konar, D.~N. Tibarewala, and M.~Hayashibe, ``A generic
  transferable eeg decoder for online detection of error potential in target
  selection,'' \emph{Frontiers in Neuroscience}, vol.~11, p. 226, 2017.

\bibitem{9177281}
F.~{Fahimi}, S.~{Dosen}, K.~K. {Ang}, N.~{Mrachacz-Kersting}, and C.~{Guan},
  ``Generative adversarial networks-based data augmentation for brain-computer
  interface,'' \emph{IEEE Transactions on Neural Networks and Learning
  Systems}, pp. 1--13, 2020.

\bibitem{data_augment}
T.~T. Um, F.~M.~J. Pfister, D.~Pichler, S.~Endo, M.~Lang, S.~Hirche,
  U.~Fietzek, and D.~Kuli\'{c}, ``Data augmentation of wearable sensor data for
  parkinson’s disease monitoring using convolutional neural networks,'' in
  \emph{Proceedings of the 19th ACM International Conference on Multimodal
  Interaction}, 2017, p. 216–220.

\end{thebibliography}
\bibliographystyle{IEEEtran}
\end{document}